\documentclass[aip, pop, reprint,numerical]{revtex4-1}

\usepackage{commands}
\usepackage{float}

\begin{document}

\title{On a variational formulation of the weakly nonlinear magnetic Rayleigh--Taylor instability}

\author{D.~E.~Ruiz}
\affiliation{Sandia National Laboratories, P.O. Box 5800, Albuquerque, New Mexico 87185-1186, USA}

\date{\today}


\begin{abstract}

The magnetic-Rayleigh—Taylor (MRT) instability is a ubiquitous phenomenon that occurs in magnetically-driven Z-pinch implosions.  It is important to understand this instability since it can decrease the performance of such implosions.  In this work, I present a theoretical model for the weakly nonlinear MRT instability.  I obtain such model by asymptotically expanding an action principle, whose Lagrangian leads to the fully nonlinear MRT equations.  After introducing a suitable choice of coordinates, I show that the theory can be cast as a Hamiltonian system, whose Hamiltonian is calculated up to sixth order in a perturbation parameter.  The resulting theory captures the harmonic generation of MRT modes.  In particular, it is shown that the saturation amplitude of the linear MRT instability grows as the stabilization effect of the magnetic-field tension increases.  Overall, the theory provides an intuitive interpretation of the weakly nonlinear MRT instability and provides a systematic approach for studying this instability in more complex settings.

\end{abstract}

\maketitle

\section{Introduction}

The magnetic-Rayleigh--Taylor instability (MRTI) always occurs in pinch plasmas in which the $\vec{J}\times \vec{B}$, or magnetomotive, force is used to compress matter.\cite{Ryutov:2000zz,Haines:2011ck,Bellan:2012wp}  In the laboratory, applications of pinch plasmas include magnetized inertial fusion,\cite{Lindemuth:2015fu,Kirkpatrick:2017kp,Lindemuth_1983,Sieman:1999um,Lindemuth:1995dh,Intrator_2002} wire-array Z-pinches,\cite{Matzen_1999,Bowers:1996gu,Pikuz:2002ca,Lebedev:2000je,Kantsyrev:2014ga} and equation-of-state studies.\cite{Lemke:2003gg}  In the particular case of magnetized inertial fusion, the MagLIF experimental platform uses high magnetic pressures acting on a cylindrical, metal liner to adiabatically compress a fuel plasma to fusion-relevant conditions.\cite{Slutz:2012gp,Slutz:2010hd,Sefkow:2014ik,Gomez:2014et}  In this fusion scheme, the MRTI grows on the liner surfaces, which can be understood as follows:  here the driving magnetic pressure plays the role of a light fluid pushing on the liner, which acts as a heavy fluid.  In analogy with the classical Rayleigh--Taylor instability (RTI), this physical configuration is dynamically unstable.  In general, it is important to better understand MRTI since it can compromise the integrity of metal liners, which in turn, is a significant factor for determining target performance in MagLIF.\cite{Knapp:2017jwb}

The magnetic-Rayleigh--Taylor (MRT) instability has been extensively studied throughout the years.  From the experimental standpoint, \Refs{Sinars:2010de,Sinars:2011ik} presented the first radiograph time sequences of seeded MRT instabilities on Z-pinch implosions.  Shortly afterwards, MRT growth was experimentally characterized on smooth coated and uncoated liners.\cite{McBride:2013gd,Awe:2016de}  In \Ref{Awe:2013dt}, the spontaneous appearance of helical structures in axially magnetized Z pinches was reported.  These structures were also investigated in further detail in \Refs{YagerElorriaga:2016eb,YagerElorriaga:2016db}.  Finally, MRTI was also experimentally studied in planar geometry in \Ref{Zier:2012fz}.  These studies have provided valuable data for benchmarking multiphysics codes and MRT theories.  

From the theoretical perspective, the first studies on MRTI were done by Kruskal and Schwarzchild,\cite{Kruskal:1954gj} Chandrasekhar,\cite{Chandrasekhar:1961uk} and Harris.\cite{Harris:1962hu}  After those seminal works, linear MRTI was further investigated by including various additional effects.  For example, \Refs{Lau:2011iw,Weis:2014kh} discussed linear MRTI in a slab geometry while including magnetic fields embedded inside the conducting fluid.  MRTI was also investigated in cylindrical geometry,\cite{Budko:1989bl} and the analysis was extended to finite-width shells.\cite{Weis:2015hk}  The stabilizing effects due to finite compressibility and elasticity of the heavy conducting fluid were reported in \Refs{Yang:2017bma,Sun:2014hq,Piriz:2019eg,Piriz:2018cx}.  Similarly, sheared flows and sheared magnetic fields were found to be MRT stabilizing in \Refs{Zhang:2005it,Zhang:2012io}.  Finally, Bell--Plesset effects in imploding shells were discussed in \Refs{Velikovich:2015jl,Schmit:2016tb}, and the effects due to finite-Larmor radius were also investigated in \Ref{Huba:1996hb}.

The theoretical studies above have primarily investigated MRTI in the linear phase.  However, it is well known that MRT perturbations can develop strong nonlinear structures during current-driven implosions.\cite{McBride:2013gd,Awe:2016de,Awe:2013dt}  In this regard, numerical simulations have been used to study nonlinear MRTI in Z-pinch implosions.  As an example, the effects of MRTI on the integrity of imploding cylindrical liners were studied in \Refs{Peterson:1996cn,Douglas:1998ca}.  The emergence of helical structures in axially magnetized Z pinches was also investigated numerically.\cite{Seyler:2018fs}  From the theoretical perspective, interesting results on nonlinear RTI and MRTI in accelerating planar slabs and cylindrical implosions have been reported.\cite{Ott:1972ct,Basko:1994co,Bashilov:1977vm,Desjarlais:1999ec,Ryutov:2014hr}  Concerning MRTI in cylindrical implosions,\cite{Basko:1994co,Bashilov:1977vm,Desjarlais:1999ec,Ryutov:2014hr} these studies used the so-called thin-shell approximation where the wavelength of the perturbations is large compared to the shell thickness.  Although this approximation only covers a subset of possible MRTI modes, it does allow to analytically investigate the fully nonlinear stages of this instability with relatively simple mathematical methods.

As a continuation of the previously mentioned studies on MRTI, here I present a theoretical model for the weakly nonlinear (wNL) MRTI in the context of the single-interface MRT problem studied in \Refs{Kruskal:1954gj,Chandrasekhar:1961uk} (see \Fig{fig:diagram}). It is worth noting that the approach presented in this work is \emph{not} based on asymptotically approximating the equations of motion for MRTI.  Instead, I construct a wNL theory for MRTI by using variational methods.  The main idea is the following.  First, I identify a variational principle for the nonlinear MRTI equations.  Second, I approximate the MRT Lagrangian by expressing the dynamical fields in terms of Fourier components and by truncating the Lagrangian up to a certain order in an asymptotic parameter.  Third, I obtain the corresponding equations of motion for wNL MRTI by varying the approximated action.  This procedure provides a systematic approach to study wNL MRTI and leads to a set of Hamiltonian equations that self-consistently conserve the energy of the system.  The resulting theory captures the harmonic generation of MRT Fourier modes and the nonlinear feedback between them.  Moreover, after obtaining asymptotic solutions for the corresponding wNL MRTI equations, I calculate the saturation amplitude of the linear MRTI, and I show that the saturation amplitude grows as the stabilization effect of the magnetic-field tension increases.  Overall, the present theory sheds light to the wNL phase of the MRT instability and can be extended for future analyses of MRTI in more complex settings, \eg for finite-width slabs or cylindrical shells.

The present work is organized as follows.  In \Sec{sec:VP}, I present a variational principle for the nonlinear MRTI equations.  In \Sec{sec:reduced}, I outline the general approach for constructing the wNL theories for MRTI.  In \Sec{sec:linear}, I approximate the exact MRT Lagrangian to its lowest order, and I recover the linear MRTI dynamics.  In \Sec{sec:nonlinear2}, I discuss the double-harmonic wNL MRT theory.  At this order of the approximation, the first two MRT Fourier harmonics are retained, and their nonlinear interaction is captured.  After obtaining the corresponding asymptotic solutions for the resulting equations, I calculate the saturation amplitude for linear MRTI.  In \Sec{sec:nonlinear3}, I briefly discuss the triple-harmonic wNL MRT theory which keeps the first three MRT Fourier harmonics.  Final conclusions and remarks on future work are given in \Sec{sec:conclusions}.  \App{app:nonlinear2} contains auxiliary calculations for the double-harmonic wNL MRT theory.

\section{Variational principle for MRT}
\label{sec:VP}

For the present study, I shall revisit the single-interface MRT problem posed by Kruskall and Schwarzchild\cite{Kruskal:1954gj} and later by Chandrasekhar.\cite{Chandrasekhar:1961uk}  I consider a semi-infinite fluid slab which I assume to be incompressible, irrotational, unmagnetized, and perfectly conducting.  The fluid is subject to a time-dependent gravitational field $\vec{g}(t)=-g(t) \vec{e}_z$ along the $\vec{e}_z$ direction, and surface tension is neglected.  The lower boundary of the fluid is described by $\xi=\xi (t,\vec{x})$, where $\vec{x} \doteq (x,y)$ is the coordinate in the $xy$ plane.  From the irrotational-flow assumption, the fluid velocity field is written as $\vec{v}(t,\vec{x},z) \doteq - \del \phi$, where $\phi=\phi(t,\vec{x},z)$ is the flow potential.  To meet the incompressibility assumption $(\del \cdot \vec{v}=0)$, the flow potential must then satisfy Laplace's equation $\del^2 \phi=0$. 

\begin{figure}
	\includegraphics[scale=.5]{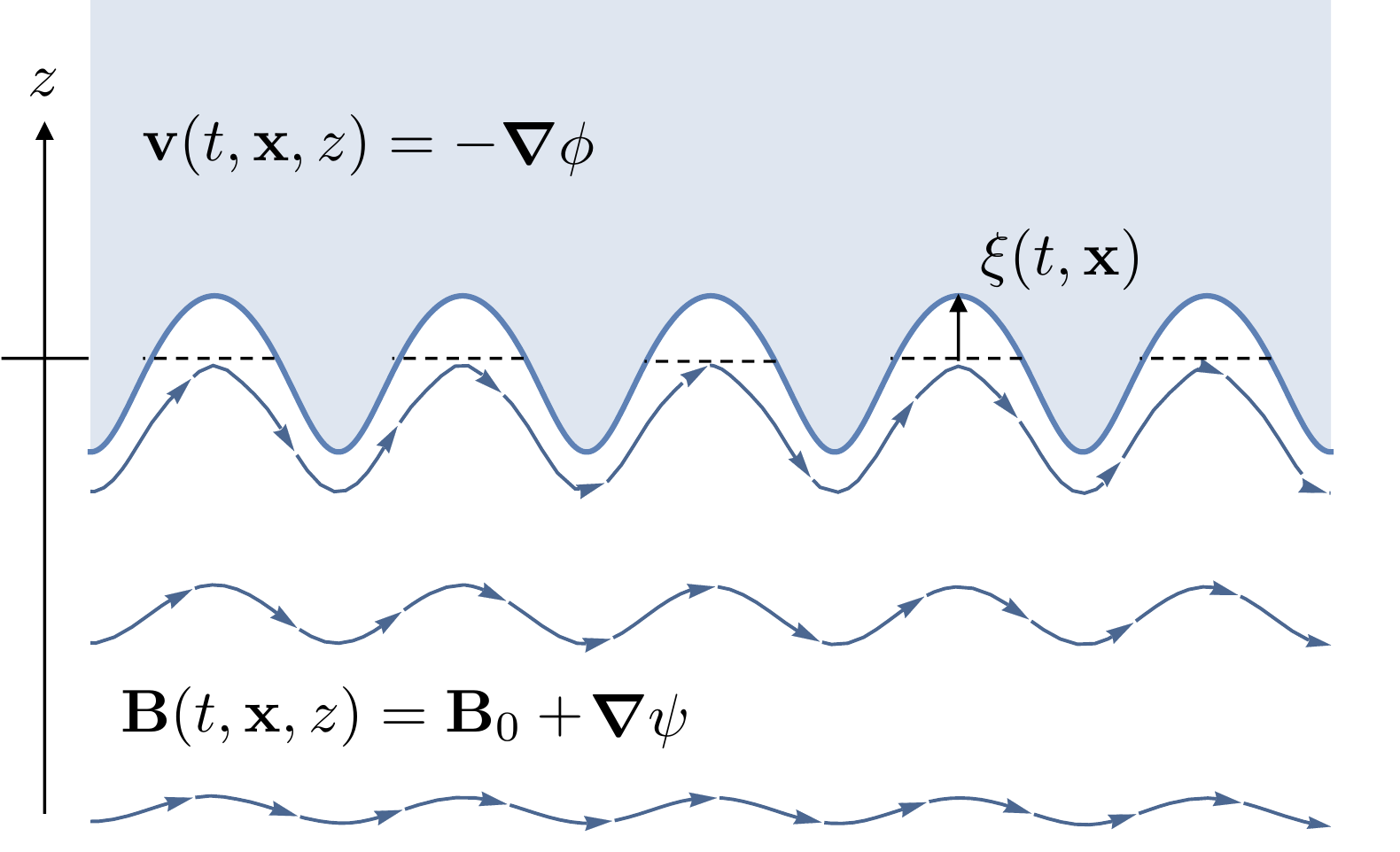}
	\caption{Schematic diagram of the geometry considered for the present MRTI study.  The fluid slab is denoted by the shaded region.  The magnetic field is represented by the stream lines.  The field $\xi(t,\vec{x})$ describes the fluid--vacuum interface.}
	\label{fig:diagram}
\end{figure}

The fluid slab interacts with a magnetic field $\vec{B}$ located in the vacuum region $(z < \xi )$.  The magnetic field satisfies $\del \cdot \vec{B}=0$ and $\del \times \vec{B}=0$.  The magnetic field is composed of a time-independent background component and a reactive component that changes in time due to the motion of the perfectly conductive fluid.  Thus, the magnetic field is written as $\vec{B}(t,\vec{x},z)=\vec{B}_0 + \vec{B}_1$, where $\vec{B}_0$ is a spatially homogeneous background magnetic field parallel to the $z=0$ plane.  To satisfy the constraints for the magnetic field, the reactive magnetic field can be written as $\vec{B}_1(t,\vec{x},z) \doteq \del \psi$, where $\psi=\psi(t,\vec{x},z)$ is the magnetic potential and satisfies Laplace's equation $\del^2 \psi=0$.  Figure \ref{fig:diagram} shows a diagram of the system considered.

One way to construct well-controlled asymptotic approximations for dynamical systems is to directly approximate a variational principle $\delta \Lambda=0$ from which the exact equations can be derived.\cite{Ruiz:2015dv,Ruiz:2017ij,Ruiz:2017et,Burby:2019vp}  Based on a well-known variational principle used in quantum hydrodynamics,\cite{Ruiz:2015bz} the action $\Lambda$ for the MRTI can be written as
\begin{equation}
	\Lambda  = \int_{t_1}^{t_2} L[\xi, \phi , \psi , \pd_t \xi,  \pd_t \phi ] \, \mathrm{d} t.
	\label{eq:basic:action}
\end{equation}
The Lagrangian $L$ of the system can be separated into a fluid component $L_{\rm fluid}$ and a magnetic component $L_{\rm B}$:
\begin{equation}
	L	\doteq	L_{\rm fluid}[\xi,\phi , \pd_t \xi,  \pd_t \phi] + L_{\rm B}[\xi, \psi],
	\label{eq:basic:L}
\end{equation}
where
\begin{align}
	L_{\rm fluid}	& \doteq
			\int _{D} \int_{\xi}^{+\infty}	\rho \left[ \frac{\partial }{\partial t}\phi	- \frac{1}{2}(\del \phi )^2 - g z \right] \mathrm{d} z \,\dx , 
	\label{eq:basic:Lfluid}\\
	L_{\rm B}		& \doteq 
			\frac{1}{8\pi }\int _{D} \int_{-\infty}^{\xi}	| \vec{B}_0+ \del  \psi |^2 \, \mathrm{d} z \,\dx .
	\label{eq:basic:LB}
\end{align}
The action \eq{eq:basic:action} is a functional of the fields $\xi$, $\phi$, and $\psi$.  In \Eq{eq:basic:Lfluid}, $\rho$ is the (constant) fluid density, and the integration domain $D$ is a $d_x \times d_y$ periodic box in the $xy$ plane.  Note that the field $\xi$ describing the fluid--vacuum interface appears in the integration boundaries in \Eqs{eq:basic:Lfluid} and \eq{eq:basic:LB}.  From a field-theoretical perspective, $\xi$ plays the role of the coupling term between the fluid motion and magnetic-field dynamics.  To more conveniently take variations of the action \eq{eq:basic:action}, I rewrite the Lagrangian components in terms of the Heaviside step function $\Theta(x)$:
\begin{align}
	L_{\rm fluid}	& =	\! \int_D \! \int_{-\infty}^\infty  	\rho \! \left[ \frac{\partial }{\partial t}\phi	- \frac{1}{2}(\del \phi )^2  - g z \right]  
								\! \Theta(z-\xi) \, \mathrm{d} z \,\dx , \\
	L_{\rm B}		& = 	\frac{1}{8\pi } \int_D \int_{-\infty}^\infty   \big| \vec{B}_0+ \del  \psi \big|^2 \, 
										\Theta(\xi-z) \, \mathrm{d} z \,\dx .
\end{align}

In the following, I verify that the action \eq{eq:basic:action} indeed leads to the nonlinear equations for MRTI.  Varying the action with respect to the flow potential $\phi$ gives
\begin{equation}
	\delta \phi \colon \quad \partial _t \Theta ( z- \xi )- \del  \cdot [ \del \phi \,  \Theta ( z- \xi ) ] =0.
\end{equation}
(Here ``$\delta \phi:$" denotes that the Euler--Lagrange equation is obtained by extremizing the action integral with respect to $\phi$.)  Explicitly calculating the derivatives leads to
\begin{equation}
	\delta ( z-  \xi ) \left( \pd _t \xi -  \del  \phi  \cdot \del  \xi  + \pd_z \phi \right) + \Theta ( z-  \xi ) \del^2\phi =0.
	\label{eq:basic:phi_variation}
\end{equation}
Thus, inside the fluid slab $[z>  \xi(t,\vec{x})]$, the flow potential satisfies Laplace's equation
\begin{equation}
	\del^2\phi  =0 .
	\label{eq:basic:phi_Laplace}
\end{equation}
At the fluid interface, where $z =  \xi(t,\vec{x})$, \Eq{eq:basic:phi_variation} leads to the nonlinear advection equation of the fluid interface:
\begin{equation}
	\big[ \pd_t \xi -  \del \phi \cdot \del \xi + \pd_z \phi \big]_{z= \xi} =0.
	\label{eq:basic:kinematic}
\end{equation}
This equation constitutes a dynamical equation for $\xi$.  Note that the gradients of the flow potential $\phi$ are evaluated at the fluid interface $z= \xi(t,\vec{x})$ in \Eq{eq:basic:kinematic}.  

In a similar manner, when varying the action with respect to the magnetic potential $\psi $, I obtain
\begin{equation}
	\delta \psi \colon \quad 
		\del \cdot  \left[ \left(\vec{B}_0 +  \del  \psi  \right) \Theta (  \xi -z) \right]=0.
\end{equation}
Since $\del \cdot \vec{B}_0 =0$, I can then write
\begin{equation}
	\delta ( \xi -z)\left(\vec{B}_0+  \del \psi \right) \cdot \del ( \xi -z)+  \Theta ( \xi -z) \del ^2\psi  =0.
\end{equation}
Thus, in the vacuum region $[z< \xi(t,\vec{x})]$, the magnetic potential satisfies Laplace's equation
\begin{equation}
	\del ^2\psi =0.
	\label{eq:basic:psi_Laplace}
\end{equation}
The magnetic field satisfies the following constraint at the boundary between the fluid and vacuum regions:
\begin{equation}
	\del ( \xi -z) \cdot \left[ \vec{B}_0+  \del \psi \right]_{z=  \xi } =0.
	\label{eq:basic:boundary}
\end{equation}
Physically, this boundary condition denotes that the magnetic field $\vec{B}$ is parallel to the surface of the perfectly conducting fluid.

Finally, when varying the action with respect to the field $\xi $ and using $\delta [\Theta(z- \xi)] = -  \delta(z-  \xi) \delta \xi$, I obtain
\begin{align}
	\delta \xi \colon \quad 
		0= & - \int_{-\infty}^\infty \rho \left[  \frac{\partial }{\pd t}\phi  - \frac{1}{2}(\del \phi )^2 - g z \right] 
				\delta ( z- \xi ) \, \mathrm{d} z   \notag \\
			 & +\frac{1}{8\pi } \int_{-\infty}^\infty \big| \vec{B}_0+  \del  \psi  \big|^2
				\delta (  \xi -z) \, \mathrm{d} z .
\end{align}
Integrating along the $z$ variable then gives
\begin{equation}
	\rho \left[  \frac{\pd }{\pd t}\phi  - \frac{1}{2}(\del \phi )^2 - g z \right]_{z=  \xi }
			=  \frac{1}{8\pi }  \big| \vec{B}_0 +  \del \psi  \big|^2_{z=  \xi } .
	\label{eq:basic:cinematic}
\end{equation}
This equation constitutes a dynamical equation for the fluid flow potential $\phi$.  As it can be shown from momentum conservation for irrotational fluids, this equation states that  the fluid pressure is equal to the magnetic pressure at the fluid--vacuum interface.  

Equations \eq{eq:basic:phi_Laplace}, \eq{eq:basic:kinematic}, \eq{eq:basic:psi_Laplace}, \eq{eq:basic:boundary}, and \eq{eq:basic:cinematic} are complemented by the periodic boundary conditions on the $xy$ plane and by the boundary conditions
\begin{gather}
	\lim_{z\to +\infty} \phi =0, \qquad 
	\lim_{z\to -\infty} \psi =0.
	\label{eq:basic:boundary_II}
\end{gather}
These equations constitute the nonlinear governing equations for the MRT instability.\cite{Kruskal:1954gj,Chandrasekhar:1961uk,Harris:1962hu}  Hence, \Eq{eq:basic:action} is a valid action for the MRT problem considered here.

For the following calculations, it will be convenient to cast the action \eq{eq:basic:action} in a form that is more reminiscent of Hamiltonian systems.  First, I integrate by parts the term in the action \eq{eq:basic:action} that involves the time derivative of the potential flow $\phi$.  The Lagrangian is then written as a sum of a symplectic part $L_{\rm sym}$ and of a Hamiltonian part $H$:
\begin{equation}
	L = L_{\rm sym}  - H.
	\label{eq:basic:L_decomposed}
\end{equation}
Here the symplectic part $L_{\rm sym}$ is given by
\begin{align}
	L_{\rm sym} 
		&	\doteq 	- \frac{2}{d_x d_y}	\int_D \int_{-\infty}^\infty 	 \phi \,	\pd_t  \Theta(z-  \xi) \, 
							\mathrm{d} z \,\dx  	\notag \\
		&	= 			\frac{2}{d_x d_y} \int_D  \Phi	\, \pd_t \xi \, \dx , 
	\label{eq:basic:L_symplectic}
\end{align}
where $\Phi=\Phi(t,\vec{x})$ is the velocity potential evaluated at the fluid interface; \ie
\begin{equation}
	\Phi(t,\vec{x}) \doteq \phi \boldsymbol{(} t,\vec{x},z=  \xi(t,\vec{x}) \boldsymbol{)}.
	\label{eq:basic:Phi}
\end{equation}
[In the above, I multiplied the Lagrangian by the constant $2/(d_x d_y \rho)$ which will be convenient later on.  The factor $d_xd_y$ is the area of the periodic box $D$.]  The Hamiltonian in \Eq{eq:basic:L_decomposed} can be decomposed as
\begin{equation}
	H \doteq H_{\rm kin} + H_{\rm g} + H_{\rm B},
	\label{eq:basic:H}
\end{equation}
where $H_{\rm kin}$, $H_{\rm g}$, and $H_{\rm B}$ are respectively the kinetic, gravitational, and magnetic components of the Hamiltonian.  They are given by
\begin{align}
	H_{\rm kin} 	& \doteq \frac{1}{d_x d_y} \int_D \int_{-\infty}^\infty 	
							(\del \phi )^2\Theta(z- \xi) \, 
							\mathrm{d} z \,	\dx, 
					\label{eq:basic:H_kinetic}	\\
	H_{\rm g} 		& \doteq \frac{2}{d_x d_y} \int_D \int_{-\infty}^\infty	
							g z \, \Theta(z- \xi) \, 
							\mathrm{d} z \, \dx, 
					\label{eq:basic:H_potential}	\\
	H_{\rm B}		& \doteq -\frac{1}{d_x d_y} \frac{1}{4 \pi \rho }	\int_D \int_{-\infty}^\infty 
							| \vec{B}_0+  \del  \psi |^2 \, \Theta( \xi-z)  \, 
							\mathrm{d} z \,	\dx .
					\label{eq:basic:H_magnetic}
\end{align}

At this point, it is worth making note of the following.  In the Lagrangian \eq{eq:basic:L_decomposed}, one should consider $\xi, \Phi$, and $\psi$ as the independent fields.  Here $\xi$ and $\Phi$ appear as a pair of canonical-conjugate variables, while $\psi$ acts as a constraint for their dynamics. Note, however, that the kinetic Hamiltonian \eq{eq:basic:H_kinetic} is still written in terms of $\phi$.  Later on, I shall invert the relation in \Eq{eq:basic:Phi} in order to write $\phi$ in terms of $\xi$ and $\Phi$.

It is also worth mentioning that, for the classical Rayleigh--Taylor problem, only the $H_{\rm kin}$ and $H_{\rm g}$ components of the Hamiltonian are kept. These terms of the Hamiltonian were previously reported by Zakharov\cite{Zakharov:1985ir,Zakharov:1972hc} for studying wNL surface waves and also used by Berning and Rubenchik\cite{Berning:1998ja} for investigating the wNL Rayleigh--Taylor and Richtmeyer--Meshkov instabilities. 

\section{Reduced variational principle for weakly nonlinear MRT}
\label{sec:reduced}

As in previous studies of the wNL RTI (see, \eg \Refs{Ingraham:1954kd,Jacobs:1988ds,Liu:2012ck,Wang:2013hl,Wang:2014dd,Wang:2015he,Guo:2017hv,Zhang:2018bw}), I shall represent the fields in terms of Fourier series.  Since $\phi$ and $\psi$ must satisfy Laplace's equation [see \Eqs{eq:basic:phi_Laplace} and \eq{eq:basic:psi_Laplace}] and the boundary conditions \eq{eq:basic:boundary}, they can be written as\cite{foot:psi}
\begin{gather}
	\phi(t,\vec{x},z) 	
			=	  	\sum_{n \in \mathbb{Z}^+ }	\ep^{n}	\F{\phi}_n(t) \cos(n \vec{k} \cdot \vec{x}) e^{-n k z}  , 
			\label{eq:reduced:Fphi}	\\
	\psi(t,\vec{x},z) 		
			=	  	\sum_{n \in \mathbb{Z}^+ }	\ep^{n}	\F{\psi}_n(t) \sin(n \vec{k} \cdot \vec{x}) e^{n k z}    .
			\label{eq:reduced:Fpsi}	
\end{gather}
Here $\F{\phi}_n$ and $\F{\psi}_m$ denote the Fourier components of $\phi$ and $\psi$.  The wavevector $\vec{k}$ has discrete allowed values $\vec{k}\doteq (2\pi m/d_x,2\pi n/d_y)$ according to the dimensions of the periodic box, and $k\doteq |\vec{k}|$.  Also, $\mathbb{Z}^+$ denotes the set of positive integers.  Similarly, the fields $\xi$ and $\Phi$ are written in the Fourier basis as follows:
\begin{gather}
	\xi(t,\vec{x}) 		=	  	\sum_{n \in \mathbb{Z}^+ }	\ep^{n} \, \F{\xi}_n(t) 		\, 	 \cos(n \vec{k} \cdot \vec{x}) , 
			\label{eq:reduced:Fxi}	\\
	\Phi(t,\vec{x})	=		\sum_{n \in \mathbb{Z}^+ }		\ep^{n} \, \F{\Phi}_n(t) 	\,	 \cos(n \vec{k} \cdot \vec{x}) .
			\label{eq:reduced:FPhi}	
\end{gather}

Note that I introduced the small parameter $\ep\ll1$ in \Eqs{eq:reduced:Fphi}--\eq{eq:reduced:FPhi} in order to explicitly denote the smallness of the MRT perturbations.  [In dimensionless variables, the parameter $\ep$ would be $\mc{O}(\xi / \lambda)$, where $\lambda$ is the characteristic wavelength of the MRT pertubations.]  This parameter will serve as the ordering parameter for the perturbation analysis that will be done in the following. 

Equations \eq{eq:reduced:Fphi}--\eq{eq:reduced:FPhi} are now substituted into the Lagrangian \eq{eq:basic:L_decomposed}.  In particular, inserting \Eqs{eq:reduced:Fxi} and \eq{eq:reduced:FPhi} into $L_{\rm sym}$ in \Eq{eq:basic:L_symplectic} gives
\begin{align}
	L_{\rm sym} 
			&	=	\frac{2}{d_x d_y} \sum_{n \in \mathbb{Z}^+ } \sum_{m \in \mathbb{Z}^+ } 
					 \int_D  \bigg[ \ep^{n+m} \, \F{\Phi}_m	\frac{\mathrm{d} \F{\xi}_n}{\mathrm{d} t}  \, 
					 \notag \\
			&			\qquad 	\qquad \times \cos( n  \vec{k} \cdot \vec{x} ) \cos( m  \vec{k} \cdot \vec{x} )  \, \bigg] \dx  
					 \notag \\
			&	=	\sum_{n \in \mathbb{Z}^+ }
					 \ep^{2n} \, \F{\Phi}_{n}	\, \frac{\mathrm{d} \F{\xi}_n }{\mathrm{d} t}  .
	\label{eq:reduced:Lsym}
\end{align}
To obtain the result above, I wrote the cosine functions in terms of exponentials and used $\int_D   \exp(i n \vec{k} \cdot \vec{x}  ) \, \dx  = d_x d_y \, \delta_{n,0}$, where $\delta_{n,m}$ is the Kronecker delta.

For the calculation of the Hamiltonian \eq{eq:basic:H}, the general procedure is to introduce \Eqs{eq:reduced:Fphi}--\eq{eq:reduced:FPhi} into \Eqs{eq:basic:H_kinetic}--\eq{eq:basic:H_magnetic} and write the Hamiltonian \emph{only} in terms of the Fourier components $\smash{\F{\xi}}_n$ and $\smash{\F{\Phi}}_n$.  This can be done by using \Eqs{eq:basic:boundary} and \eq{eq:basic:Phi} to write $\smash{\F{\phi}}_n$ and $\smash{\F{\psi}}_n$ in terms of $\smash{\F{\xi}}_n$ and $\smash{\F{\Phi}}_n$.  After doing so, the resulting Lagrangian for the system will be of the following form:
\begin{equation}
	L	=	\sum_{n \in \mathbb{Z}^+ } 
						\bigg(	\ep^{2n} \, \F{\Phi}_n	\, \frac{\mathrm{d}  \F{\xi}_n  }{\mathrm{d} t}	\bigg)
					 - H(t, \F{\xi},\F{\Phi} ),
	\label{eq:reduced:L}
\end{equation} 
where $\F{\xi}$ and $\F{\Phi}$ denote the sets of Fourier coefficients appearing in \Eqs{eq:reduced:Fxi} and \eq{eq:reduced:FPhi}.  In Secs.~\ref{sec:linear}--\ref{sec:nonlinear3}, I will explicitly calculate the Hamiltonian $\smash{H(t,\F{\xi},\F{\Phi})}$ in the Fourier representation.  For now, it is only  important to note that the accuracy (in $\ep)$ of the proposed wNL MRT theory depends only on the order in $\ep$ to which the Hamiltonian $H$ is calculated.

Hamilton's equations for $\smash{\F{\xi}_n}$ and $\smash{\F{\Phi}_n}$ are obtained by varying \Eq{eq:reduced:L} with respect to $\smash{\F{\xi}_n}$ and $\smash{\F{\Phi}_n}$.  The resulting equations, which are valid to all orders in $\ep$, are
\begin{align}
	\delta \F{\Phi}_{n}: 	\quad	\frac{\mathrm{d} \F{\xi}_n }{\mathrm{d} t}	
				& =	\frac{1}{\ep^{2n}} 	\frac{\pd H}{\pd\F{\Phi}_n}	,	
	\label{eq:reduced:xi_n}		\\
	\delta \F{\xi}_{n}: 		\quad	\frac{\mathrm{d} \F{\Phi}_n }{\mathrm{d} t} 	
				& =	-\frac{1}{\ep^{2n}}		\frac{\pd H}{\pd \F{\xi}_n} .
	\label{eq:reduced:Phi_n}	
\end{align}

\section{Single-harmonic linear MRTI}
\label{sec:linear}

To obtain the linear approximation of MRTI, it is sufficient to calculate the Hamiltonian $H$ up to $\mc{O}(\ep^2)$.  Regarding $H_{\rm kin}$ in \Eq{eq:basic:H_kinetic}, integrating by parts leads to
\begin{equation}
	H_{\rm kin} 	=	 -  \frac{1}{d_x d_y} \int_D \int_{-\infty}^{\infty}
								\phi \del \cdot [ (\del \phi )\Theta(z-\xi) ] \, \mathrm{d} z \,\dx,
	\label{eq:linear:Hkin_aux}	
\end{equation} 
where I used the periodic boundary conditions and \Eqs{eq:basic:boundary_II}.  Since I shall later substitute the Fourier representation of $\phi$ into $H_{\rm kin}$, I can use the fact that $\phi$ satisfies Laplace's equation.  Integrating along $z$ gives
\begin{equation}
	H_{\rm kin} 	= -  \frac{1}{d_x d_y} \int_D 	\Phi  (\del \phi )_{z= \xi} \cdot \del( z- \xi) \, \dx,
	\label{eq:linear:Hkin}	
\end{equation} 
where $\Phi$ is defined in \Eq{eq:basic:Phi}. Upon substituting  \Eqs{eq:reduced:Fphi} and \eq{eq:reduced:FPhi} into the above, I obtain $H_{\rm kin}  	= (\ep^2 k /2) \smash{ \F{\phi}_1 \F{\Phi}_1} +\mc{O}(\ep^3)$. I can then eliminate $\smash{\F{\phi}_1}$ by using $\smash{\F{\phi}_1 \simeq \F{\Phi}_1}$, which is obtained by inserting \Eqs{eq:reduced:Fphi} and \eq{eq:reduced:FPhi} into \Eq{eq:basic:Phi} and linearizing.  To lowest order, the kinetic Hamiltonian is approximated by
\begin{equation}
	H_{\rm kin}  	= \ep^2 \frac{k}{2} \F{\Phi}_1^2+\mc{O}(\ep^3).
	\label{eq:linear:Hkin_approx}
\end{equation}

In a similar manner, one can calculate the gravitational Hamiltonian \eq{eq:basic:H_potential}.  Integrating along $z$ and substituting \Eq{eq:reduced:Fxi} leads to
\begin{equation}
	H_{\rm g} 	= - \ep^2 \frac{g}{2} \, \F{\xi}_1^2 +\mc{O}(\ep^4),
	\label{eq:linear:Hg}
\end{equation}
where I dropped a constant (infinite) term since it does not affect the equations of motion.

To calculate $H_{\rm B}$ in \Eq{eq:basic:H_magnetic}, I first separate the squared norm of the magnetic field into its different components and integrate by parts.  Since the magnetic potential satisfies Laplace's equation, I obtain
\begin{align}
	H_{\rm B} 	=  & -	\frac{1}{d_x d_y} \frac{1}{4\pi \rho }	\int_D \int^{\xi}_{-\infty} | \vec{B}_0 |^2 \,  \mathrm{d} z \,	\dx 
									\notag \\
							&+	\frac{1}{d_x d_y}  \frac{1}{2\pi \rho }	\int_D  \psi \vec{B}_0 \cdot \del [\Theta(\xi-z)]  \, \mathrm{d} z \,	\dx 
									\notag \\
							&+	\frac{1}{d_x d_y}  \frac{1}{4\pi \rho }	\int_D \psi \del{\psi} \cdot \del [\Theta(\xi-z)]   \, \mathrm{d} z \,	\dx ,
	\label{eq:linear:HB_aux}
\end{align} 
where I used $\del \cdot \vec{B}_0 =0$ in the above.  After integrating the first term along $z$, I obtain a constant (infinite) term (which can be dropped) and a term that is linear to $\xi$.  This last term disappears after integrating on the domain $D$.\cite{foot:cylindrical}  After taking the gradient of the Heaviside step function, integrating along $z$, and using \Eq{eq:basic:boundary}, I then obtain
\begin{equation}
	H_{\rm B} 	=  	\frac{1}{d_x d_y}  \frac{1}{4\pi \rho }	\int_D  \psi \big|_{z=\xi} \, \vec{B}_0 \cdot \del \xi  \,	\dx  .
	\label{eq:linear:HB}
\end{equation}
Thus, the magnetic Hamiltonian is expressed as the integral on the $xy$ plane of the magnetic potential $\psi$ evaluated at the perturbation surface multiplied by the gradient of $\xi$ along the background magnetic field $\vec{B}_0$.  As a reminder, \Eq{eq:linear:HB} is only valid when $\psi$ satisfies \Eq{eq:basic:boundary}.  In terms of Fourier components, \Eq{eq:linear:HB} is written as
\begin{equation}
 	H_{\rm B} 	=  -	\ep^2 \frac{\vec{k}\cdot \vec{B}_0}{8\pi \rho }	\F{\psi}_{1} \F{\xi}_1 + \mc{O}(\ep^3).
 	\label{eq:linear:HB_aux_II}
\end{equation}
The Fourier component $\F{\psi}_{1}$ must now be written in terms of $\F{\xi}_1$.  When linearizing \Eq{eq:basic:boundary} and substituting \Eqs{eq:reduced:Fpsi} and \eq{eq:reduced:Fxi}, it can be shown that $\smash{\F{\psi}_1 \simeq - k^{-1} (\vec{k}\cdot \vec{B}_0 ) \F{\xi}_1}+\mc{O}(\ep^3)$.	  Inserting this expression into \Eq{eq:linear:HB_aux_II} gives
\begin{equation}
	H_{\rm B} 	=  	\ep^2 \frac{(\vec{k}\cdot \vec{v}_A)^2}{2k} \, \F{\xi}_1^2 +\mc{O}(\ep^3),
	\label{eq:linear:HB_approx}
\end{equation}
where $\vec{v}_A \doteq \vec{B}_0 / \sqrt{4\pi \rho}$ is the Alfvén velocity.

In summary, after collecting the results in \Eqs{eq:linear:Hkin_approx}, \eq{eq:linear:Hg}, and \eq{eq:linear:HB_approx}, I obtain the following Lagrangian for the single-mode linear MRTI:
\begin{equation}
	L = \ep^2 \, \F{\Phi}_1 \, \frac{\mathrm{d}\F{\xi}_1}{\mathrm{d} t} 
			- H(t, \F{\xi}_1, \F{\Phi}_1 ) +\mc{O}(\ep^3),
	\label{eq:linear:Lagr}
\end{equation}
where the Hamiltonian $H$ is given by
\begin{equation}
	H = 	\ep^2 \frac{k}{2} \F{\Phi}_1^2 
			-  \frac{\ep^2}{2} \left(  g - \frac{(\vec{k}\cdot \vec{v}_A)^2}{k}  \right)  \F{\xi}_{1}^2 . 
	\label{eq:linear:Ham}
\end{equation}
In analogy to the classical phase-space Lagrangian for point particles $\smash{L= \vec{P} \cdot \dot{\vec{X}} - H(t,\vec{X},\vec{P})}$, $\smash{\F{\xi}_1}$ and $\smash{\F{\Phi}_{1}}$ play the roles of a generalized coordinate and of a canonical momentum, respectively.  In the Hamiltonian \eq{eq:linear:Ham}, the term $\smash{k \F{\Phi}_1^2/2}$ can be interpreted as the kinetic energy of the system, where $m_k \doteq 1/k$ is the mass of the MRT ``point particle."  From this perspective, the fact that RTI modes grow faster for higher $k$ modes is because higher $k$ modes are less massive and thus accelerate faster.  The second term in \Eq{eq:linear:Ham} represents a quadratic potential.  When $k g > (\vec{k}\cdot \vec{v}_A)^2$, the potential energy is negative, thus leading to unstable behavior.  In the opposite case when $k g < (\vec{k}\cdot \vec{v}_A)^2$, the potential energy is positive, and the temporal dynamics will be oscillatory in nature.  Thus, the stabilizing behavior of the external magnetic field on MRTI is recuperated.\cite{Kruskal:1954gj,Chandrasekhar:1961uk,Harris:1962hu}

\begin{figure*}
	\includegraphics[scale=.44]{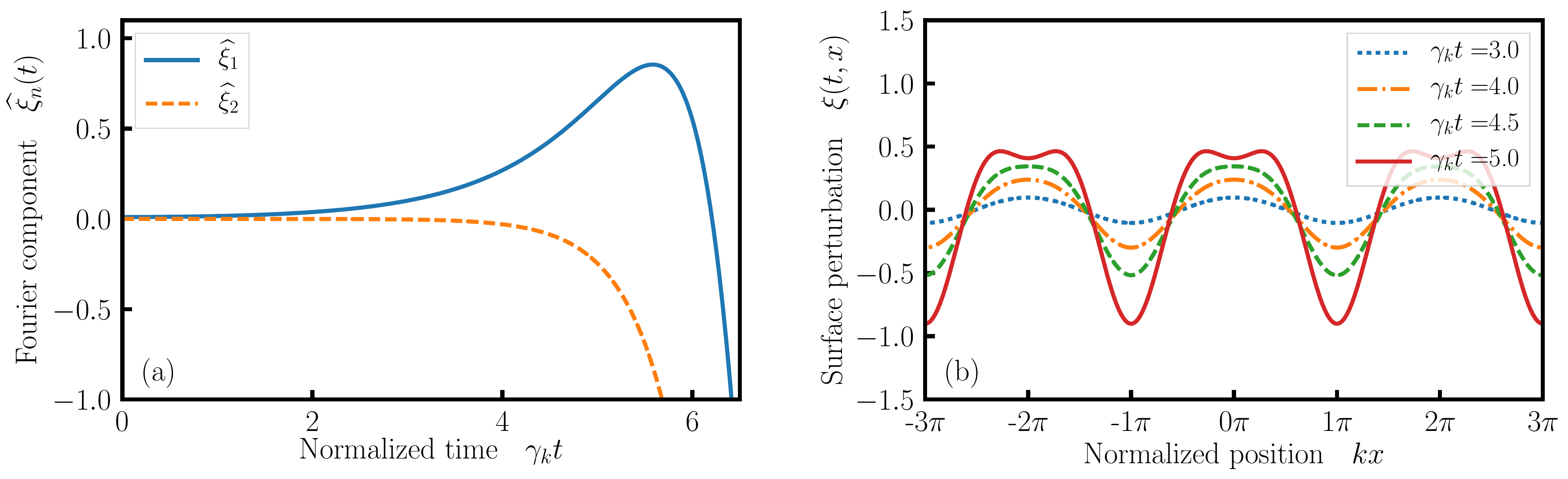}
	\caption{(a) Time evolution of the first and second harmonics of MRTI.  (b) Surface perturbation $\xi(t,x)$ evaluated at different times $\gamma_\vec{k}t=\{3,4,4.5,5.0\}$.  The problem considered is one dimensional, where $x$ is the spatial coordinate.  The initial conditions are $\smash{\F{\xi}_1(0)=0.01}$, $\smash{\F{\xi}_2(0)=0}$, and $\smash{\F{\Phi}_{1,2}(0)=0}$.  The fundamental wavenumber is $k=1$, the gravity constant is $g=1$, and the perturbation parameter is set to $\ep=1$.  No external magnetic field was considered so $\sigma =(\vec{k}\cdot \vec{v}_A)^2/(kg) = 0$.}
	\label{fig:nonlinear2:no_B}
\end{figure*}

The Hamiltonian equations generated by the Lagrangian \eq{eq:linear:Lagr} lead to the linear MRT equations:
\begin{align}
	\delta \F{\Phi}_1: 	\quad 	
			\frac{\mathrm{d} \F{\xi}_1 }{\mathrm{d} t}  		&	= k \F{\Phi}_1,  
		\label{eq:linear:xi1}	\\
	\delta \F{\xi}_1: 		\quad 	
			\frac{\mathrm{d} \F{\Phi}_1}{\mathrm{d} t}  	&	= \left( g - \frac{ \left(\vec{k} \cdot \vec{v}_A \right)^2}{k} \right) \F{\xi}_1.
		\label{eq:linear:Phi1}
\end{align}
Combining these two equations leads to the well-known equation for the linear MRTI surface perturbation:
\begin{equation}
	\frac{\mathrm{d}^2 \F{\xi}_1}{\mathrm{d} t^2} 
		- \gamma_\vec{k}^2(t) \, \F{\xi}_1 =0,
	\label{eq:linear:MRT}
\end{equation}
where
\begin{equation}
	\gamma_\vec{k}(t) \doteq [ kg(t) -  (\vec{k} \cdot \vec{v}_A )^2 ]^{1/2}
	\label{eq:linear:gamma}
\end{equation}
is the instantaneous MRT linear growth rate.\cite{Kruskal:1954gj,Chandrasekhar:1961uk,Harris:1962hu}  Note that it is convenient to write $\gamma_\vec{k}(t)$ as
\begin{equation}
	\gamma_\vec{k}(t) = \sqrt{k g } \, \sqrt{1-\sigma (t,\vec{k})},
\end{equation}
where $\sqrt{k g }$ is the growth rate for classical RTI and $\sigma(t, \vec{k}) \doteq (\vec{k}\cdot \vec{v}_A)^2/[kg(t)]$ is the ratio between the stabilizing magnetic-tension and the gravitational acceleration.

For initial conditions where $\xi(0,\vec{x}) = \smash{\F{\xi}_1(0) \cos( \vec{k}\cdot \vec{x})}$ and the fluid is at rest, the linear solution is $\smash{\F{\xi}_1(t) = \F{\xi}_1(0) \cosh( \gamma_\vec{k} t)}$ for time-independent $g$. Thus, the time evolution of the surface perturbation is given by
\begin{equation}
	\xi(t,\vec{x}) = \F{\xi}_1(0) \cosh( \gamma_\vec{k} t) \cos( \vec{k}\cdot \vec{x}).
	\label{eq:linear:surface}
\end{equation}

As a final comment, note that one can also obtain a variational principle that directly leads to \Eq{eq:linear:MRT}.  From \Eq{eq:linear:Phi1}, one can write $\smash{\F{\Phi}_1 = k^{-1} \mathrm{d}_t \F{\xi}_1}$.  Substituting this into the phase-space Lagrangian \eq{eq:linear:Lagr} then leads to
\begin{equation}
	L( \F{\xi}_1, \dot{\F{\xi}}_1) =  \frac{\ep^2}{2k} \left( \frac{\mathrm{d} \F{\xi}_1 }{\mathrm{d} t}\,   \right)^2 +  \frac{\ep^2}{2} \frac{\gamma_\vec{k}^2(t)}{k}  \, \F{\xi}_{1}^2.
\end{equation}
It is simple to verify that varying the action with respect to $\F{\xi}_1$ gives \Eq{eq:linear:MRT}.

\section{Double-harmonic weakly-nonlinear MRTI}
\label{sec:nonlinear2}

To obtain a wNL MRT theory that includes the interaction between the first and second MRT harmonics, one needs to calculate the Hamiltonian up to $\mc{O}(\ep^4)$.  In this case, the Lagrangian of the system is given by
\begin{align}
	L = & \sum_{n=1}^2	\bigg( \ep^{2n} \, \F{\Phi}_{n}  \frac{\mathrm{d}   \F{\xi}_n }{\mathrm{d} t}  \bigg)
		  	 - H(t, \F{\xi}_1, \F{\xi}_2 , \F{\Phi}_1, \F{\Phi}_2) 
			+	\mc{O}(\ep^5),
		\label{eq:nonlinear2:Lagr}	
\end{align}
where the Hamiltonian is
\begin{align}
	 H	 = &	\, \sum_{n=1}^2 \ep^{2n} \left( \frac{nk}{2} \F{\Phi}_n^2 - \frac{\gamma_{n \vec{k}}^2}{2nk}  \, \F{\xi}_n^2 \right)
	 			\notag \\
	 		& - \ep^4 \frac{k^3 }{8} \,  \F{\xi}_1^2   \, \F{\Phi}_1^2 
				+ \ep^4 \frac{k^2}{2} \, \F{\xi}_2  \, \F{\Phi}_1^2 
				\notag \\
			&	-  \frac{\ep^4}{2} |\vec{k}\cdot \vec{v}_A|^2 \, \F{\xi}_1^2 \F{\xi}_2
				-  \ep^4 \frac{k}{8} |\vec{k}\cdot \vec{v}_A|^2 \, \F{\xi}_1^4.
	\label{eq:nonlinear2:Ham}	
\end{align}
(Details on the calculations of the Hamiltonian $H$ are included in Appendix \ref{app:nonlinear2}.)  The equations of motion for the first and second Fourier coefficients are obtained by varying the action with the Lagrangian \eq{eq:nonlinear2:Lagr}.  This gives
\begin{align}
	\delta \F{\Phi}_1: \quad
		\frac{\mathrm{d}  \F{\xi}_1 }{\mathrm{d} t}
			& =	k\F{\Phi}_1
					-	\ep^2 \frac{k^3}{4}  \F{\Phi}_1 \F{\xi}_1^2
					+ \ep^2 k^2 \F{\Phi}_1 \F{\xi}_{2}  ,		
		\label{eq:reduced:ELE_xi_1}	\\
	\delta \F{\Phi}_2 : \quad
		\frac{\mathrm{d} \F{\xi}_2 }{\mathrm{d} t} 
			&	 = 2k \F{\Phi}_2,
		\label{eq:reduced:ELE_xi_2} \\
	\delta \F{\xi}_1: \quad
		\frac{\mathrm{d} \F{\Phi}_1 }{\mathrm{d} t} 
			&  =	\frac{\gamma^2_\vec{k}(t)}{k} \F{\xi}_1 
					+\ep^2 \frac{k^3}{4} \F{\xi}_1 \F{\Phi}_1^2
					+ \ep^2  |\vec{k}\cdot \vec{v}_A|^2 \F{\xi}_1 \F{\xi}_2
					\notag \\
			&	\quad
					+ \ep^2 \frac{k}{2} (\vec{k}\cdot \vec{v}_A)^2 \F{\xi}_1^3 ,	
		\label{eq:reduced:ELE_Phi_1}	 \\
	\delta \F{\xi}_2 : \quad
		\frac{\mathrm{d}\F{\Phi}_2 }{\mathrm{d} t} 
			& 	= 	\frac{\gamma^2_{2\vec{k}}(t)}{2k} \F{\xi}_2 
					- \frac{k^2}{2} \F{\Phi}_1^2
					+ \frac{1}{2} \, (\vec{k}\cdot \vec{v}_A)^2 \F{\xi}_1^2, 
		\label{eq:reduced:ELE_Phi_2} 
\end{align}
These are the governing equations for the double-harmonic wNL MRTI.

\begin{figure*}
	\includegraphics[scale=.44]{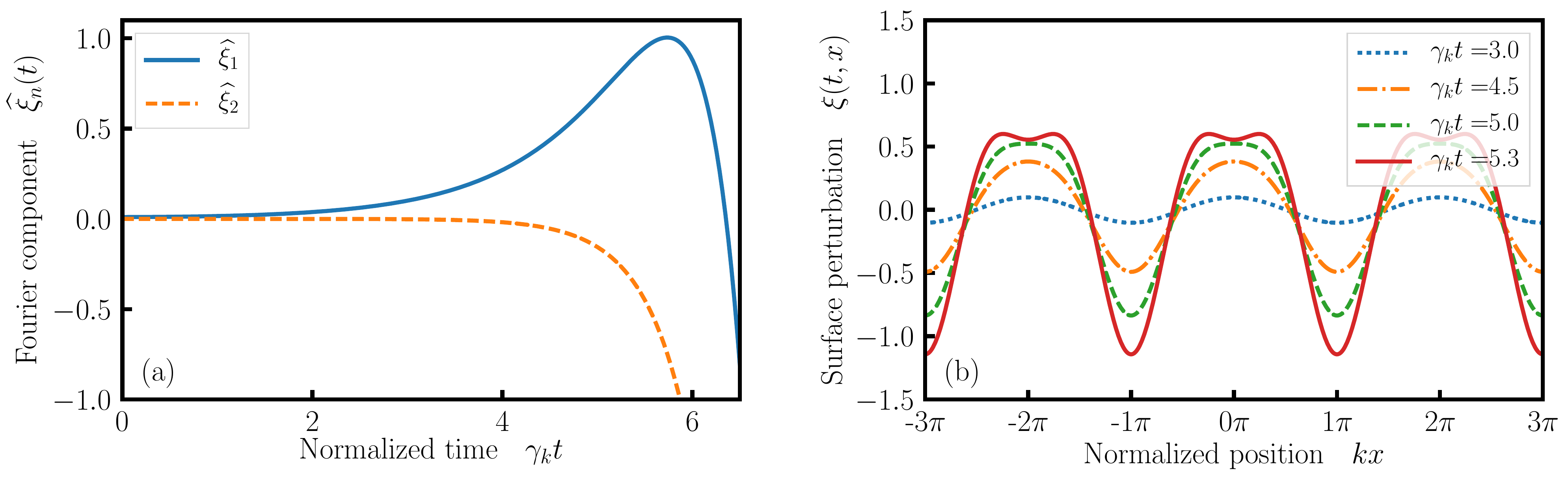}
	\caption{(a) Time evolution of the first and second harmonics of the MRT instability.  (b) Surface perturbation $\xi(t,x)$ evaluated at different times.  Same initial conditions and parameters were used as in \Fig{fig:nonlinear2:no_B} with the exception that the external magnetic field was chosen so that $\sigma  =(\vec{k}\cdot \vec{v}_A)^2/(kg) = 0.2$.}
	\label{fig:nonlinear2:B}
\end{figure*}

To discuss the temporal dynamics described by \Eqs{eq:reduced:ELE_xi_1}--\eq{eq:reduced:ELE_Phi_2}, it is perhaps more instructive to comment on the Hamiltonian $H$ rather than on the equations themselves. When one varies the action, the terms inside the sum in \Eq{eq:nonlinear2:Ham} lead to the linear terms appearing in \Eqs{eq:reduced:ELE_xi_1}--\eq{eq:reduced:ELE_Phi_2}.  The second row in \Eq{eq:nonlinear2:Ham} contains nonlinear coupling terms originating from the kinetic Hamiltonian $H_{\rm kin}$.  The first term proportional to $\smash{\F{\Phi}_1^2 \, \F{\xi}_1^2 }$ represents a nonlinear self-coupling of the first harmonic.  The second term containing $\smash{\F{\Phi}_1^2 \, \F{\xi}_2}$ describes a coupling between the first and second MRT harmonics.  For the case of classical RTI, the latter is, in fact, responsible for the nonlinear driving of the second harmonic by the first harmonic.  The third row in \Eq{eq:nonlinear2:Ham} contains nonlinear coupling terms of magnetic origin.  Similarly to before, the first term proportional to $\smash{\F{\xi}_1^2 \F{\xi}_2}$ represents a coupling between the two MRTI modes, and the second term containing $\smash{\F{\xi}_1^4}$ represents a nonlinear self-coupling of the first harmonic.  It is worth noting that, contrary to the lowest-order contribution of the magnetic energy (which is stabilizing), the magnetic self-coupling term proportional to $\smash{\F{\xi}_1^4}$ appears to be MRT destabilizing. 

Based on the previous discussion, it is not surprising that \Eqs{eq:reduced:ELE_xi_1}--\eq{eq:reduced:ELE_Phi_2} already include the nonlinear forcing of the second harmonic $\smash{\F{\Phi}_2}$, as well as the nonlinear feedback on the first Fourier coefficients $\smash{\F{\xi}_1}$ and $\smash{\F{\Phi}_1}$.  To obtain these effects using the traditional approach for building wNL theories, one has to expand the exact equations of motion up to third order in $\ep$.  (See, \eg \Refs{Ingraham:1954kd,Jacobs:1988ds,Liu:2012ck,Wang:2013hl,Wang:2014dd,Wang:2015he,Guo:2017hv,Zhang:2018bw} for applications of wNL theory to RTI.)  This approach also gives equations for the third Fourier coefficients $\smash{\F{\xi}_3}$ and $\smash{\F{\Phi}_3}$.  In contrast, in the procedure presented in this paper, the third harmonics do not yet appear at this order in the theory.  However, this wNL MRTI model retains the Hamiltonian property of the parent model and, in consequence, \emph{conserves} an asymptotic approximation of the energy of the system.

To evaluate the effects of the external magnetic field on the wNL MRT growth, I numerically solved \Eqs{eq:reduced:ELE_xi_1}--\eq{eq:reduced:ELE_Phi_2} using a fourth-order Runge--Kutta algorithm.  I first discuss the classical RTI case when no background magnetic field is present.  Figure \ref{fig:nonlinear2:no_B} presents the temporal evolution of the Fourier coefficients $\smash{\F{\xi}_1}$ and $\smash{\F{\xi}_2}$ and of the surface perturbation $\xi(t,x)$.  As shown in \Fig{fig:nonlinear2:no_B}{\color{blue} (a)}, during the first one or two $e$-folding times, the amplitude of the first MRTI mode is small, and $\smash{\F{\xi}_1}$ grows exponentially.  As the fundamental MRTI mode becomes sufficiently strong, it eventually begins to drive the second MRTI harmonic.  When the nonlinear self-coupling and coupling with the second harmonic are no longer negligible, the growth of $\smash{\F{\xi}_1}$ saturates near $\gamma_\vec{k}t \simeq 5.0$.  Afterwards, $\smash{\F{\xi}_1}$ reaches a maximum value and then rapidly decreases.  The resulting temporal evolution of the surface perturbation is shown in \Fig{fig:nonlinear2:no_B}{\color{blue} (b)}.  As expected, one observes the formation of bubbles and spikes on the surface perturbation.  However, for $\gamma_\vec{k}t \gtrsim 5.0$, the rounding of the bubbles begins to deform.  This behavior is not physical and signals the breakdown of wNL theory.\cite{Berning:1998ja}  Interestingly, the time $\gamma_\vec{k}t \simeq 5.0$ roughly coincides with the saturation of the growth rate of $\smash{\F{\xi}_1}$.

Figure \ref{fig:nonlinear2:B} presents the numerical solution of \Eqs{eq:reduced:ELE_xi_1}--\eq{eq:reduced:ELE_Phi_2} when an external magnetic field is present.  When comparing Figures \ref{fig:nonlinear2:no_B}{\color{blue} (a)} and \ref{fig:nonlinear2:B}{\color{blue} (a)}, one observes that the Fourier coefficients $\smash{\F{\xi}_1}$ and $\smash{\F{\xi}_2}$ follow similar behavior when plotted against the number of $e$-folding times $\gamma_\vec{k}t $.  Of course, the MRTI modes are stabilized by the magnetic-field tension and grow more slowly in real time units.  When carefully comparing the temporal evolution of $\smash{\F{\xi}_1}$ in \Figs{fig:nonlinear2:no_B}{\color{blue} (a)} and \ref{fig:nonlinear2:B}{\color{blue} (a)}, one observes that $\smash{\F{\xi}_1}$ has a slightly larger value for the second case at the time of maximum growth.  The peak-to-valley amplitude of the surface perturbation $\xi(t,x)$ in \Fig{fig:nonlinear2:B}{\color{blue} (b)} also appears to be larger at the time of bubble deformation when compared to \Fig{fig:nonlinear2:no_B}{\color{blue} (b)}.  As I shall discuss next, this is related to an increase of the saturation amplitude of the linear MRTI due to the stabilizing effect of the magnetic-field tension.

It is instructive to analytically calculate the temporal behavior of the solutions of \Eqs{eq:reduced:ELE_xi_1}--\eq{eq:reduced:ELE_Phi_2} far from the transient phase but before the breakdown of wNL theory.  For this analysis, one can use the methods of iterative solutions\cite{Wang:2013hl,Wang:2014dd,Jacobs:1988ds,Liu:2012ck,Wang:2015he,Zhang:2018bw,Guo:2017hv,Ingraham:1954kd,Haan:1991jg} and of dominant balance.\cite{Bender:1999ee}  I look for solutions in the following form:
\begin{equation}	
	\begin{aligned}
	\F{\xi}_{1} & = \F{\xi}_{1}^{(0)} + \ep^2 \F{\xi}_{1}^{(1)} +..., &
	\F{\xi}_2 &= \F{\xi}_{2}^{(0)} +... , \\
	\F{\Phi}_1 & = \F{\Phi}_{1}^{(0)} + \ep^2 \F{\Phi}_{1}^{(1)} +  ... , &
	\F{\Phi}_{2} &= \F{\Phi}_{2}^{(0)} + ...
	\end{aligned}
	\label{eq:nonlinear2:asymptotic_expansion}
\end{equation}
For same initial conditions, the lowest-order terms $\F{\xi}_{1}^{(0)}$ and $\smash{\F{\Phi}_{1}^{(0)}}$ correspond to the linear solutions of MRTI discussed in \Sec{sec:linear}.  Since I am interested in calculating the long-time behavior of the solutions, I can keep the asymptotic dominant terms only.  This gives
\begin{align}
	\F{\xi}_{1}^{(0)} &
		= \F{\xi}_{1}(0) \cosh( \gamma_\vec{k} t) 
		\sim \frac{1}{2} \, \F{\xi}_{1}(0) \exp(\gamma_\vec{k} t), 
		\label{eq:nonlinear2:xi_10}	\\
	\F{\Phi}_{1}^{(0)} &
		= \frac{\gamma_\vec{k}}{k} \, \F{\xi}_{1}(0) \sinh( \gamma_\vec{k} t) 
		\sim \frac{\gamma_\vec{k}}{2k} \, \F{\xi}_{1}(0) \exp(\gamma_\vec{k} t).
		\label{eq:nonlinear2:Phi_10}
\end{align}

\begin{figure*}
	\includegraphics[scale=.56]{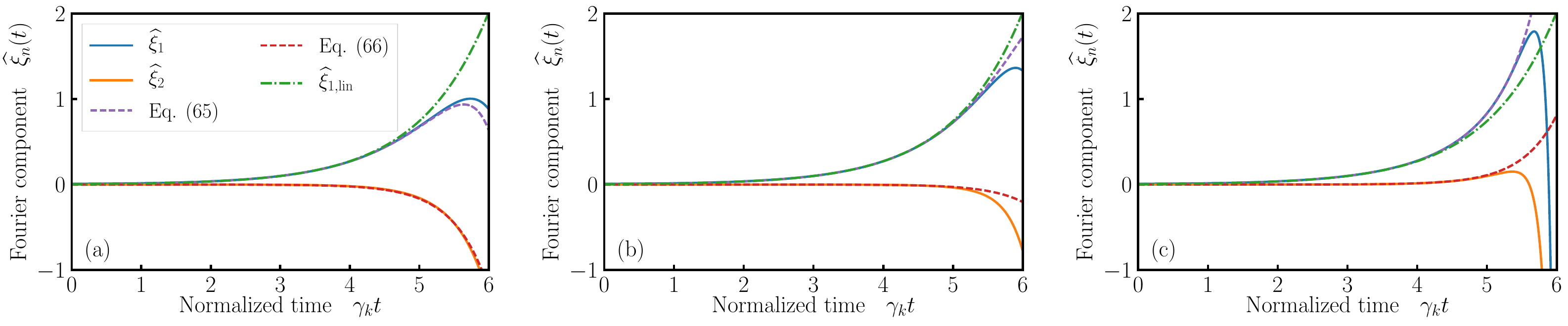}
	\caption{Comparison between the asymptotic expressions in \Eqs{eq:nonlinear2:xi1_asymptotic}--\eq{eq:nonlinear2:xi2_asymptotic} and the numerical solutions of \Eqs{eq:reduced:ELE_xi_1}--\eq{eq:reduced:ELE_Phi_2}.  Same initial conditions and parameters were used as in \Fig{fig:nonlinear2:no_B}.  The subfigures correspond to $\sigma = 0.2$ with $\sigma =\{0.20, 0.45,0.70\}$.}
	\label{fig:nonlinear_asymptotic}
\end{figure*}

I now insert \Eqs{eq:nonlinear2:xi_10} and \eq{eq:nonlinear2:Phi_10} into \Eqs{eq:reduced:ELE_xi_2} and \eq{eq:reduced:ELE_Phi_2}.  At sufficiently large times, the dominant terms of the solutions for $\smash{\F{\xi}_2}$ and $\smash{\F{\Phi}_2}$ will be proportional to $\exp(2\gamma_\vec{k} t)$.\cite{foot:exp}  Hence, I look for solutions of the form $\smash{\F{\xi}_2^{(0)} \sim A \exp(2\gamma_\vec{k} t)}$ and $\smash{\F{\Phi}_2^{(0)} \sim B \exp(2\gamma_\vec{k} t)}$, where the coefficients $A$ and $B$ are to be determined.  This leads to the following algebraic system of equations:
\begin{equation}
	\begin{pmatrix}
		2\gamma_\vec{k} & - 2k \\
		-\gamma^2_{2\vec{k}}/ (2k) & 2 \gamma_\vec{k} 
	\end{pmatrix}
	\begin{pmatrix}
		A \\ B
	\end{pmatrix}
	=
	- \frac{[ \F{\xi}_{1}(0) ]^2}{16} 
	\begin{pmatrix}
		0 \\ \gamma^2_{2\vec{k}}
	\end{pmatrix}.
\end{equation}
Inverting the matrix above yields
\begin{gather}
	A	=	- \frac{\gamma^2_{2\vec{k}}}{16g } [\F{\xi}_{1}(0) ]^2 , \qquad
	B	=	- \frac{\gamma_{\vec{k}} \gamma^2_{2\vec{k}}}{16gk } [ \F{\xi}_{1}(0) ]^2 .
	\label{eq:nonlinear2:xiPhi_2}
\end{gather}

The next step is to determine the dominant components of the next order terms $\smash{\F{\xi}_1^{(1)}}$ and $\smash{\F{\Phi}_1^{(1)}}$ of the first harmonic.  I substitute the results obtained in \Eqs{eq:nonlinear2:xi_10}--\eq{eq:nonlinear2:xiPhi_2} into \Eqs{eq:reduced:ELE_xi_1} and \eq{eq:reduced:ELE_Phi_1}.  As before, at large times, the dominant components of $\smash{\F{\xi}_1^{(1)}}$ and $\smash{\F{\Phi}_1^{(1)}}$ will be of the form $\smash{\F{\xi}_1^{(1)} \sim C \exp(3\gamma_\vec{k} t)}$ and $\smash{\F{\Phi}_1^{(1)} \sim D \exp(3\gamma_\vec{k} t)}$, where $C$ and $D$ are to be determined.  This leads to
\begin{equation}
	\begin{pmatrix}
		3\gamma_\vec{k} & - k \\
		-\gamma^2_{\vec{k}}/k & 3 \gamma_\vec{k} 
	\end{pmatrix}
	\begin{pmatrix}
		C \\ D
	\end{pmatrix}
	= \frac{ [ \F{\xi}_{1}(0) ]^3 }{32g} 
	\begin{pmatrix}
		-k \gamma_\vec{k}  (kg + \gamma^2_{2\vec{k}} )  \\
		gk\gamma_\vec{k}^2 + 4 (\vec{k}\cdot\vec{v}_A)^4
	\end{pmatrix}.
	\label{eq:nonlinear2:xiPhi_1correction}
\end{equation}
Solving for the coefficients $C$ and $D$ gives
\begin{align}
	C &	= 	- \frac{k(kg +  \gamma_\vec{k}^2)  \gamma_{2\vec{k}}^2 }{128g \gamma_\vec{k}^2}  \,
				[ \F{\xi}_{1}(0) ]^3 , \\
	D &	=	\frac{(\vec{k}\cdot \vec{v}_A)^2 [kg +  (\vec{k}\cdot \vec{v}_A)^2] }{64g \gamma_\vec{k}}  \,
				[ \F{\xi}_{1}(0) ]^3 .
	\label{eq:nonlinear2:asymptotic_final}
\end{align}
Upon gathering the results in \Eqs{eq:nonlinear2:xi_10}--\eq{eq:nonlinear2:asymptotic_final}, I obtain
\begin{align}
	\F{\xi}_1 & \sim \F{\xi}_{\rm 1, lin}(t)
							\left( 1 -\ep^2 \frac{k(kg +  \gamma_\vec{k}^2)  \gamma_{2\vec{k}}^2 }
									{16g \gamma_\vec{k}^2} \,  [ \F{\xi}_{\rm 1, lin}(t) ]^2 \right),	
		\label{eq:nonlinear2:xi1_asymptotic}	\\
	\F{\xi}_2 & \sim -\frac{\gamma^2_{2\vec{k}}}{4g } \,	 [ \F{\xi}_{\rm 1, lin}(t)]^2,
		\label{eq:nonlinear2:xi2_asymptotic}
\end{align}
where $\F{\xi}_{\rm 1, lin}(t) \doteq (1/2) \F{\xi}_{1}(0) \exp(\gamma_\vec{k} t)$ is the dominant component of the linear solution of $\F{\xi}_1$.

Regarding the obtained asymptotic solutions \eq{eq:nonlinear2:xi1_asymptotic} and \eq{eq:nonlinear2:xi2_asymptotic}, it is important to note that the analysis above is only valid when the fundamental MRT mode is unstable, \ie $\gamma_{\vec{k}}^2 >0$.  Otherwise, it would not have been possible to preemptively choose the asymptotic forms of the solutions.  In the case of classical RTI when no magnetic fields are present, \Eqs{eq:nonlinear2:xi1_asymptotic} and \eq{eq:nonlinear2:xi2_asymptotic} simplify to
\begin{align}
	\F{\xi}_1(t) & \sim \F{\xi}_{\rm 1, lin}(t)
							\left( 1 -\ep^2 \frac{k^2}{4}   [ \F{\xi}_{\rm 1, lin}(t) ]^2 \right),	 
		\label{eq:nonlinear2:xi1_asymptotic_RTI}	 \\
	\F{\xi}_2(t) & \sim - \frac{k}{2}  \,	 [ \F{\xi}_{\rm 1, lin}(t)]^2.
		\label{eq:nonlinear2:xi2_asymptotic_RTI}
\end{align}
These expressions agree with previous reported results for classical RTI.\cite{Ingraham:1954kd,Jacobs:1988ds}

A difference to highlight between RTI and MRTI is the following.  In the case of RTI, the Fourier component $\smash{\F{\xi}_2}$ and the $\mc{O}(\ep^2)$ correction to $\smash{\F{\xi}_1}$ are always negative [see \Eqs{eq:nonlinear2:xi1_asymptotic_RTI} and \eq{eq:nonlinear2:xi2_asymptotic_RTI}].  The $\mc{O}(\ep^2)$ correction to $\smash{\F{\xi}_1}$ is in fact responsible for the saturation of the growth rate of $\smash{\F{\xi}_1}$ observed in \Fig{fig:nonlinear2:no_B}{\color{blue} (a)}.  For the MRTI case, this is not always true: these terms are proportional to $\gamma^2_{2\vec{k}}$ [see \Eqs{eq:nonlinear2:xi1_asymptotic} and \eq{eq:nonlinear2:xi2_asymptotic}], which can change sign when the magnetic-field tension becomes sufficiently strong.  More specifically, this occurs when $ 1/2 <  \sigma < 1$, where $\sigma  \doteq (\vec{k}\cdot \vec{v}_A)^2/(kg)$.  In such regimes, $\smash{\F{\xi}_1}$ can grow faster than the linear approximation $\smash{\F{\xi}_{\rm 1, lin}}$.

In \Fig{fig:nonlinear_asymptotic}, the asymptotic expressions obtained in \Eqs{eq:nonlinear2:xi1_asymptotic} and \eq{eq:nonlinear2:xi2_asymptotic} are compared to the numerical solutions of \Eqs{eq:reduced:ELE_xi_1}--\eq{eq:reduced:ELE_Phi_2} for three different values of the parameter $\sigma$.  In all cases, the asymptotic expressions approximate well the numerical solutions for $1\lesssim \gamma_{\vec{k}} t \lesssim 5.0$; \ie in the temporal window after the transient phase and before the breakdown of wNL theory. In particular, \Fig{fig:nonlinear_asymptotic}{\color{blue} (b)} shows the case for $\sigma\simeq1/2$ where the growth rate $\gamma^2_{2\vec{k}}$ for the second MRTI harmonic tends to zero.  As expected, the second Fourier component $\smash{\F{\xi}_2}$ lies close to zero, and the temporal evolution of $\smash{\F{\xi}_1}$ follows the linear result quite closely.  As shown in  \Fig{fig:nonlinear_asymptotic}{\color{blue} (c)}, for the case where $1/2<\sigma<1$, the numerical solution for $\smash{\F{\xi}_1}$ and its asymptotic approximation indeed grow faster than the linear approximation $\smash{\F{\xi}_{\rm 1,lin}}$, which confirms the remark given previously. 

\begin{figure}[H]
	\includegraphics[scale=.43]{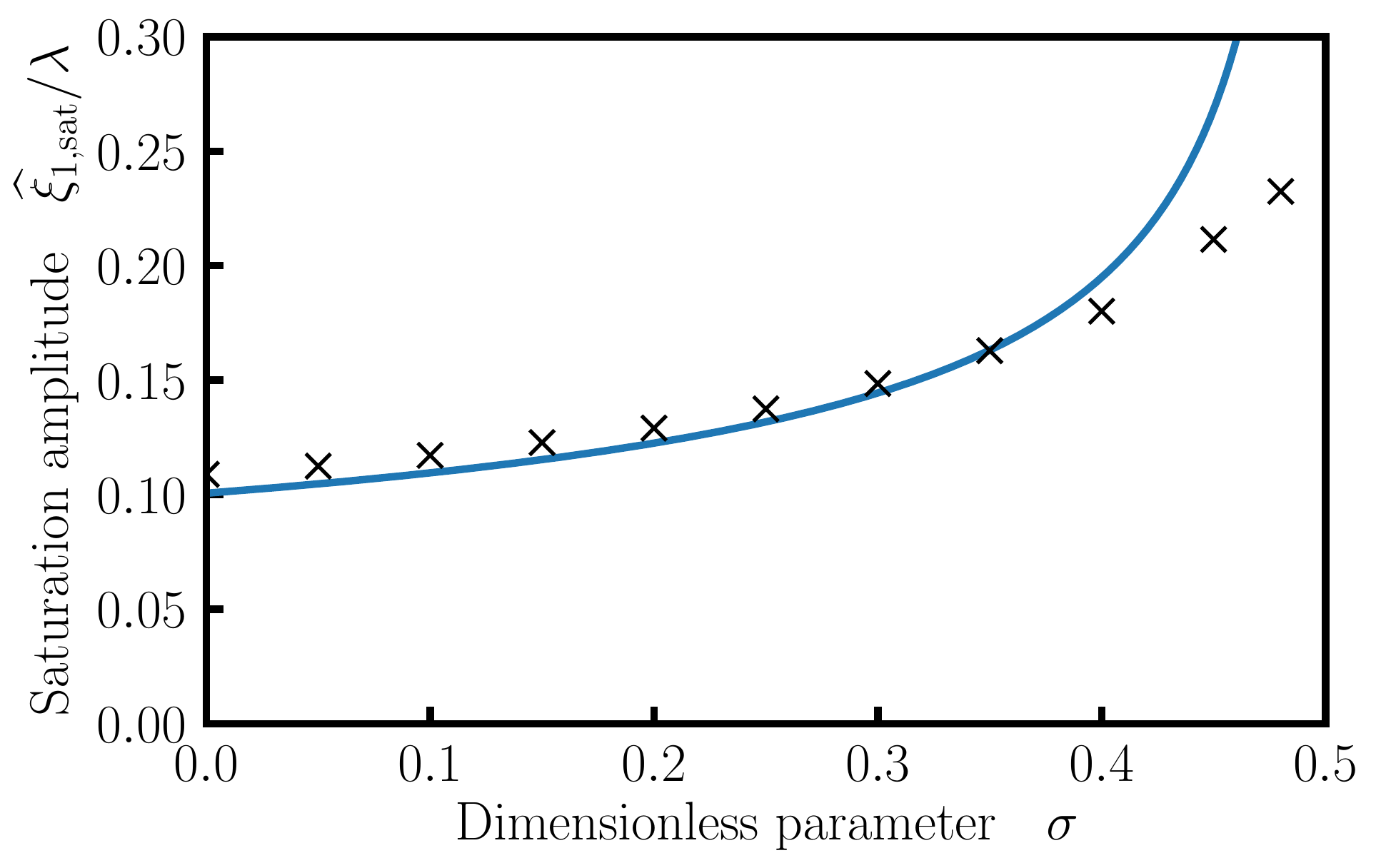}
	\caption{Saturation amplitude of linear MRTI as a function of the parameter $\sigma$.  The continuous line is given by \Eq{eq:nonlinear2:xi1_saturation}. The scatter points were obtained via numerical simulations.}
	\label{fig:function_f}
\end{figure}

In the regime of validity of \Eqs{eq:nonlinear2:xi1_asymptotic} and \eq{eq:nonlinear2:xi2_asymptotic}, one can estimate the saturation amplitude (SA) of the linear-growth phase of the first harmonic $\smash{\F{\xi}_1}$.\cite{Jacobs:1988ds}  The SA for the fundamental mode can be defined as the amplitude $\smash{\F{\xi}_{\rm 1,sat}}$ when the fundamental mode is reduced by $10\%$ in comparison to the linear solution, \ie when $\smash{(\F{\xi}_{\rm 1,lin}-\F{\xi}_1 )/\F{\xi}_{\rm 1,lin} = 0.1 }$.\cite{Atzeni:2009phys,Liu:2012ck}  From this definition, I find
\begin{equation}
	\frac{\F{\xi}_{\rm 1,sat}}{\lambda}
		=	\frac{1}{\pi} \sqrt{\frac{1}{10}} \, f(\sigma).
	\label{eq:nonlinear2:xi1_saturation}
\end{equation}
where
\begin{equation}
	f(\sigma ) \doteq \sqrt{\frac{1-\sigma }{(1-2\sigma )(1-\sigma /2)}}.
	\label{eq:nonlinear2:f}
\end{equation}

The SA in \Eq{eq:nonlinear2:xi1_saturation} is plotted as a function of the dimensionless parameter $\sigma$ in \Fig{fig:function_f}.  In the interval $0\leq \sigma \lesssim 0.4$, the calculated SA shows good agreement with the SA obtained via numerical solutions of \Eqs{eq:reduced:ELE_xi_1}--\eq{eq:reduced:ELE_Phi_2}.  In the classical RTI limit where $\sigma \simeq 0$, $\smash{\F{\xi}_{\rm 1, sat}} \simeq 0.1 \lambda$ which agrees with previously reported results.\cite{Atzeni:2009phys,Liu:2012ck,Jacobs:1988ds}  Remarkably, the SA increases as the magnetic-field tension (and hence $\sigma$) becomes larger.  Thus, although magnetic-field tension stabilizes the linear growth of MRTI, it can also increase the SA at which the linear MRTI transitions to the nonlinear phase.  This result partially explains the differences observed in \Figs{fig:nonlinear2:no_B} and \ref{fig:nonlinear2:B}.  It is to be noted that a similar effect was reported previously where considering surface tension also leads to an increase in the SA for classical RTI.\cite{Guo:2017hv}  Finally, as shown in \Eq{eq:nonlinear2:f}, $f(\sigma)$ diverges at $\sigma =1/2$.  This occurs because the $\mc{O}(\ep^2)$ correction term in \Eq{eq:nonlinear2:xi1_asymptotic} tends to zero near this limit.  To fix this issue, one would have to calculate higher-order corrections for the present wNL MRT theory.  However, doing so would lead to corrections that go beyond the accuracy of theory \eq{eq:nonlinear2:Lagr}.

\section{Triple-harmonic weakly-nonlinear MRTI}
\label{sec:nonlinear3}

The wNL MRT theory can be extended to include the first, second, and third MRT harmonics.  To calculate the corresponding $\mc{O}(\ep^6)$-accurate Lagrangian, one can use the \textsc{Mathematica} software package.\cite{Wolfram:2003wa}  For the sake of brevity, here I only report the end result:

\begin{equation}
	L =  \sum_{n=1}^3
				\left( \! \ep^{2n} \, \F{\Phi}_{n} \frac{\mathrm{d}  \F{\xi}_n }{\mathrm{d} t}  \!\right)
		  	 - H(t, \F{\xi}_1, \F{\xi}_2 ,  \F{\xi}_3 ,  \F{\Phi}_1, \F{\Phi}_2 , \F{\Phi}_3)  
			+	\mc{O}(\ep^7),
		\label{eq:nonlinear3:Lagr}	
\end{equation}
where the corresponding Hamiltonian is given by
\begin{widetext}
	\begin{align}
	 H	 = &	\, \sum_{n=1}^3 \ep^{2n} \left( \frac{nk}{2} \F{\Phi}_n^2 - \frac{\gamma_{n \vec{k}}^2}{2nk}  \, \F{\xi}_n^2 \right)
	 			- \ep^4 \frac{k^3 }{8} \,  \F{\xi}_1^2  \, \F{\Phi}_1^2 
				+ \ep^4 \frac{k^2}{2} \,  \F{\xi}_2  \, \F{\Phi}_1^2 
				-  \frac{\ep^4}{2} (\vec{k}\cdot \vec{v}_A)^2 \, \F{\xi}_1^2 \F{\xi}_2
				-  \ep^4 \frac{k}{8} (\vec{k}\cdot \vec{v}_A)^2 \, \F{\xi}_1^4
				\notag \\
			&	+2\ep^6 k^2 \F{\xi}_3  \F{\Phi}_1 \F{\Phi}_2
					+\ep^6 \frac{k^3}{4} \left( \F{\xi}_2^2 \F{\Phi}_1^2 +  \F{\xi}_1 \F{\xi}_3 \F{\Phi}_1^2 -4 \F{\xi}_1 \F{\xi}_2 \F{\Phi}_1 \F{\Phi}_2 \right)
				    +\ep^6 \frac{k^4}{4} \left( \frac{1}{3} \F{\xi}_1^3 \F{\Phi}_1 \F{\Phi}_2 - \frac{3}{2}  \F{\xi}_1^2 \F{\xi}_2 \F{\Phi}_1^2 \right)
				    + \ep^6 \frac{11k^5}{192} \F{\xi}_1^4 \F{\Phi}_1^2
				 \notag \\
			&	-2\ep^6 (\vec{k}\cdot \vec{v}_A)^2  \F{\xi}_1 \F{\xi}_2 \F{\xi}_3
				+ \ep^6\frac{k }{4}(\vec{k}\cdot \vec{v}_A)^2 \left( \F{\xi}_1^3 \F{\xi}_3 - 3 \F{\xi}_1^2 \F{\xi}_2^2 \right) 
				+ \ep^6 \frac{7 k^2}{24} (\vec{k}\cdot \vec{v}_A)^2 \F{\xi}_1^4 \F{\xi}_2
				+ \ep^6 \frac{11 k^3}{192} (\vec{k}\cdot \vec{v}_A)^2 \F{\xi}_1^6.
	\label{eq:nonlinear3:Ham}	
 	\end{align}
\end{widetext}

\begin{figure*}
	\includegraphics[scale=.44]{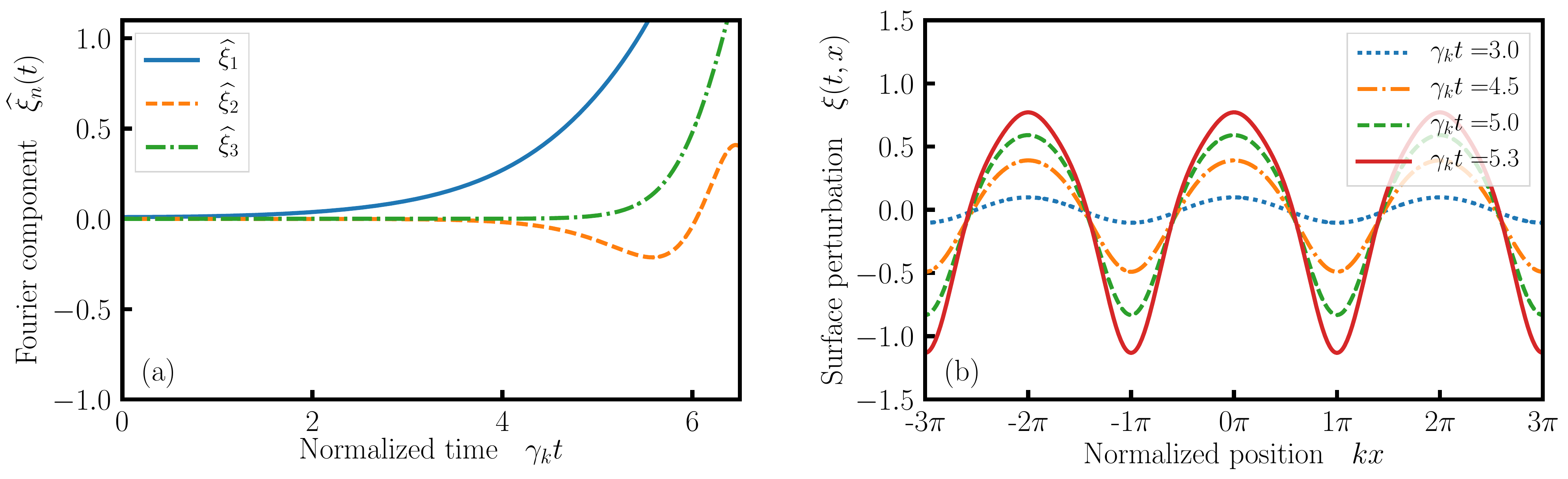}
	\caption{Time evolution of the MRT instability using the triple-harmonic weakly nonlinear model.  (a) Time evolution of the Fourier harmonics $\smash{\F{\xi}_n(t)}$.  (b) Surface perturbation $\xi(t,x)$ evaluated at different times.  Same initial conditions and parameters were used as in \Fig{fig:nonlinear2:B}.}
	\label{fig:nonlinear3:B}
\end{figure*}

The terms appearing in \Eq{eq:nonlinear3:Ham} can be interpreted as follows.  In the first line, the terms appearing in the sum lead to the linear driving terms of each Fourier harmonic.  The subsequent $\mc{O}(\ep^4)$ terms are those previously reported in \Eq{eq:nonlinear2:Ham} and were discussed in \Sec{sec:nonlinear2}.  The $\mc{O}(\ep^6)$ terms appearing in the second line of \Eq{eq:nonlinear3:Ham} are nonlinear coupling terms originating from the kinetic part of the Hamiltonian.  There, one can identify terms corresponding to nonlinear couplings between the first, second, and third harmonics, as well as higher-order interactions between the first and second harmonics and a nonlinear self-coupling of the first harmonic.  Finally, the third line in \Eq{eq:nonlinear3:Ham} includes the $\mc{O}(\ep^6)$ nonlinear coupling terms arising from the magnetic Hamiltonian.

The equations of motion of this triple-harmonic wNL MRT theory can be obtained by using \Eqs{eq:reduced:xi_n} and \eq{eq:reduced:Phi_n}.  For the sake of conciseness, I shall not write the equations for the first and second harmonics.  The corresponding equations for the third Fourier harmonic are
\begin{align}
	\delta \F{\Phi}_3: \quad
		\frac{\mathrm{d}\F{\xi}_3  }{\mathrm{d} t} 
			& =	3k\F{\Phi}_3 ,
	\label{eq:nonlinear3:ELE_xi_3}	 \\
	\delta \F{\xi}_3: \quad
		\frac{\mathrm{d} \F{\Phi}_3 }{\mathrm{d} t} 
			&  =	\frac{\gamma^2_{3\vec{k}}(t)}{3k} \F{\xi}_3 
					-2 k^2 \F{\Phi}_1  \F{\Phi}_2
					- \frac{ k^3}{4} \F{\xi}_1  \F{\Phi}_1^2
					\notag \\
			& 	\quad
					+ 2  (\vec{k}\cdot \vec{v}_A)^2 \F{\xi}_1 \F{\xi}_2
					- \frac{k}{4}  (\vec{k}\cdot \vec{v}_A)^2  \F{\xi}_1^3.
		\label{eq:nonlinear3:ELE_Phi_3}	
\end{align}
From \Eq{eq:nonlinear3:ELE_Phi_3}, one can observe that the nonlinear driving terms for $\smash{\F{\Phi}_3}$ are $\smash{-2 k^2 \F{\Phi}_1  \F{\Phi}_2}$ and $\smash{k^3 \F{\xi}_1  \F{\Phi}_1^2/4}$.  These terms are obtained from the expressions proportional to $\smash{\F{\xi}_3  \F{\Phi}_1 \F{\Phi}_2}$ and $\smash{\F{\xi}_1 \F{\xi}_3 \F{\Phi}_1^2}$ in the Hamiltonian \eq{eq:nonlinear3:Ham}.  In order to conserve energy, these terms in the Hamiltonian will also generate nonlinear feedback terms in the equations for the first and second harmonics that include the effects of $\smash{\F{\xi}_3}$.  Similar arguments apply for the nonlinear magnetic driving terms in \Eq{eq:nonlinear3:ELE_Phi_3}.

As an example, \Fig{fig:nonlinear3:B} shows the time evolution of the MRTI using the triple-harmonic wNL MRTI theory \eq{eq:nonlinear3:Lagr}.  Compared to \Fig{fig:nonlinear2:B}{\color{blue} (a)}, similar dynamics for the first and second MRT harmonics are obtained up to $\gamma_\vec{k}t\simeq 5.2$ with a small growth in the third Fourier mode $\smash{\F{\xi}_3}$.  For later times, the dynamics predicted by the two models significantly diverge, which is most likely due to the breakdown of wNL theory and lack of convergence. When comparing Figures \ref{fig:nonlinear2:B}{\color{blue} (b)} and \ref{fig:nonlinear3:B}{\color{blue} (b)}, one can observe that the third Fourier harmonic $\smash{\F{\xi}_3}$ significantly changes the shape of the MRT spikes and bubbles even when its magnitude is small.  The third Fourier harmonic $\smash{\F{\xi}_3}$ apparently fixes the roundness of the bubbles but will also eventually destroy it once it starts growing rapidly.

\section{Conclusions and future work}
\label{sec:conclusions}

In this work, I proposed a theoretical model to describe the weakly nonlinear (wNL) stage of the magnetic-Rayleigh--Taylor (MRT) instability.  I obtained the model by asymptotically expanding an exact action principle that leads to the nonlinear MRT equations.  The theory can be cast as a Hamiltonian system,  whose Hamiltonian was calculated up to sixth order in the perturbation parameter.  The obtained wNL theory describes the harmonic generation of MRT modes.  From the obtained equations, I found that the saturation amplitude of the linear MRT instability increases as the stabilizing effect of the magnetic-field tension increases.  

The present work can be extended to study the MRT instability in more complex settings.  As an example, the action principle \eq{eq:basic:action} can be modified to investigate the MRT instability in finite-width planar slabs or cylindrical shells with finite thickness.  More specifically, to study the MRT instability in planar slabs, one can introduce a new field $\eta(t,\vec{x})$ describing the perturbation of the fluid upper surface and replace the upper integration boundary in \Eq{eq:basic:Lfluid} with $a+\eta$, where $a$ is the slab width.  In principle, modifications to include an additional magnetic field in the second vacuum region or to treat the cylindrical problem could be easily done.  For the latter, it would be interesting to investigate if the experimental observations on the MRT instability reported in \Refs{Sinars:2010de,Sinars:2011ik,Awe:2013dt} can be explained using a simple wNL MRT model for a finite-thickness cylindrical shell.  This will be investigated in future works.

The author is indebted to D.~A.~Yager-Elorriaga, E.~P.~Yu, J.~R.~Fein, K.~J.~Peterson, P.~F.~Schmit, R.~A.~Vesey, C.~A.~Jennings, and M.~R.~Weis for fruitful discussions.  Sandia National Laboratories is a multimission laboratory managed and operated by National Technology $\&$ Engineering Solutions of Sandia, LLC, a wholly owned subsidiary of Honeywell International Inc., for the U.S. Department of Energy (DOE) National Nuclear Security Administration under contract DE-NA0003525.  This paper describes objective technical results and analysis. Any subjective views or opinions that might be expressed in the paper do not necessarily represent the views of the U.S. DOE or the United States Government.

\appendix

\section{Auxiliary calculations for the double-harmonic weakly-nonlinear MRT theory}
\label{app:nonlinear2}

\subsection{Calculation of the kinetic Hamiltonian $\boldsymbol{H_{\rm kin}}$}
\label{app:nonlinear2:Hkin}

To calculate $H_{\rm kin}$ in \Eq{eq:basic:H_kinetic}, I first write $\F{\phi}_n$ in terms of $\F{\xi}_n$ and $\F{\Phi}_n$.  Taylor expanding \Eq{eq:basic:Phi} leads to
\begin{equation}
	\Phi(t,\vec{x})=		\phi|_{z=0}
							+ 	\xi \, \frac{\pd \phi }{\pd z} \bigg|_{z=0} 
							+ \frac{1}{2} \xi^2\frac{\pd^2 \phi }{\pd z^2} \bigg|_{z=0} + ...
	\label{eq:nonlinear2:Phi_expanded}
\end{equation}
I now substitute \Eqs{eq:reduced:Fphi}, \eq{eq:reduced:Fxi}, and \eq{eq:reduced:FPhi} into \Eq{eq:nonlinear2:Phi_expanded}.  The equations for the first two harmonics are
\begin{align}
	\F{\Phi}_1 
			& =	 \F{\phi}_1 - \ep^2 \frac{k}{2} (2  \F{\xi}_1 \F{\phi}_2   +\F{\xi}_2  \F{\phi}_1  )
					+ \ep^2  \frac{3k^2}{8} \F{\xi}_1^2 \F{\phi}_1    +... ,	\\
	\F{\Phi}_2 
			& =	 \F{\phi}_2 -  \frac{k}{2} \F{\xi}_{1}  \F{\phi}_1  + ...
\end{align}
To solve the equations above, I use the following asymptotic ansatz:
\begin{equation}
	\F{\phi}_n = \F{\phi}_n^{(0)} + \ep^2 \F{\phi}_n^{(1)}+...
	\label{eq:nonlinear2:phi_asymptotic}
\end{equation}
To lowest order in $\ep$, I obtain
\begin{equation}
	\F{\phi}_1^{(0)} = \F{\Phi}_1 ,
	\qquad 
	\F{\phi}_2^{(0)} = \F{\Phi}_2 + \frac{k}{2} \,  \F{\Phi}_1  \F{\xi}_{1}.
	\label{eq:nonlinear2:phi_aux_I}
\end{equation}
For the second harmonic $\F{\phi}_2$, it is not necessary to compute further terms in the asymptotic series \eq{eq:nonlinear2:phi_asymptotic} because they are not necessary for calculating the Lagrangian up to $\mc{O}(\ep^4)$ in accuracy.  The $\mc{O}(\ep^2)$ correction to $\smash{\F{\phi}_1}$ is
\begin{align}
	\F{\phi}_1^{(2)} 
		&	= \frac{k}{2} \big(2    \F{\xi}_1 \F{\phi}_2^{(0)} + \F{\xi}_2 \F{\phi}_1^{(0)}   \big)
					- \frac{3 k^2 }{8}   \F{\xi}_{1}^2 \F{\phi}_1^{(0)}  
				\notag \\
		&	= 	 k   \F{\xi}_1  \F{\Phi}_2 
				+ \frac{k}{2} \F{\Phi}_1  \F{\xi}_2
				+ \frac{k^2}{8} \F{\Phi}_1 \F{\xi}_1^2	,
	\label{eq:nonlinear2:phi_aux_II}
\end{align}
where I substituted \Eqs{eq:nonlinear2:phi_aux_I}.

After having obtained the asymptotic expressions for $\smash{\F{\phi}_1}$ and $\smash{\F{\phi}_2}$, I now proceed with the explicit calculation of $H_{\rm kin}$ in \Eq{eq:basic:H_kinetic}.  Substituting \Eqs{eq:reduced:Fphi}, \eq{eq:reduced:Fxi} and \eq{eq:reduced:FPhi} into \Eq{eq:linear:Hkin} leads to
\begin{widetext}
\begin{align}
	H_{\rm kin}	
			=		 	&	\, \frac{1}{d_x d_y} 
													\sum_{n\in \mathbb{Z}^+} \sum_{m \in \mathbb{Z}^+}
													\ep^{n+m}	\int  \dx \,
													m k \left( 1 - m k \xi + \frac{m^2}{2} k^2 \xi^2 +... \right) \,
													\F{\Phi}_n \F{\phi}_m  \,
													\cos(n \vec{k}\cdot \vec{x}) \cos(m \vec{k}\cdot \vec{x}) 
													\notag \\
						&	- \frac{1}{d_x d_y} 
													\sum_{n\in \mathbb{Z}^+} 
													\sum_{m\in \mathbb{Z}^+}\
													\sum_{l\in \mathbb{Z}^+}
													\ep^{n+m+l}	
													\int  \dx \, m l k^2 \left( 1 - m k \xi  + ... \right) \,
													\F{\Phi}_n \F{\phi}_m \F{\xi}_l \,
													\cos(n \vec{k}\cdot \vec{x}) \sin(m \vec{k}\cdot \vec{x}) \sin(l \vec{k}\cdot \vec{x})  	.
	\label{app:nonlinear2:H_expansion}				
\end{align}
\end{widetext}
I now expand $H_{\rm kin}$ in an asymptotic series in $\ep$:
\begin{equation}
	H_{\rm kin}	= \ep^2 	H_{\rm kin,2} + \ep^3 H_{\rm kin,3} + \ep^4 H_{\rm kin,4} +.... 	
	\label{app:nonlinear2:H_asymptotic}			
\end{equation}
I then substitute \Eqs{eq:reduced:Fxi} and \eq{eq:nonlinear2:phi_asymptotic} into \Eq{app:nonlinear2:H_expansion} and calculate $H_{\rm kin}$ to each order in $\ep$.  At $\mc{O}(\ep^2)$, the only contribution to $H_{\rm kin,2}$ is given by
\begin{align}
	H_{\rm kin,2}	
		=	&	\, \frac{k}{2} \, \F{\Phi}_1 \F{\phi}_1^{(0)} 
		=		\, \frac{k}{2} \, \F{\Phi}_1^2,
	\label{eq:nonlinear2:Hkin2}
\end{align}
which agrees with \Eq{eq:linear:Hkin_approx}.  A contribution of $\mc{O}(\ep^3)$ to the kinetic Hamiltonian would involve a product of three fundamental Fourier harmonics.  However, such terms proportional to $\cos(\vec{k}\cdot \vec{x})\sin^2(\vec{k}\cdot \vec{x})$ would vanish when integrating on the $xy$ plane.  Thus, $H_{\rm kin,3}=0$.  Finally, the $\mc{O}(\ep^4)$ contribution is
\begin{align}
		H_{\rm kin,4}	
			=	&	\,	k \F{\Phi}_2 \F{\phi}_2^{(0)} 
						+ \frac{k}{2 }\F{\Phi}_1 \F{\phi}_1^{(1)} 
						+ \frac{k^2}{4}	\F{\Phi}_1 \F{\phi}_1^{(0)} \F{\xi}_2
						\notag \\
				& \,	- \frac{k^2}{2}	\F{\Phi}_1 \F{\phi}_2^{(0)} \F{\xi}_1
						- \frac{k^2}{2}	\F{\Phi}_2 \F{\phi}_1^{(0)} \F{\xi}_1
						+ \frac{k^3}{16} \, \F{\Phi}_1 \F{\phi}_1^{(0)} \F{\xi}_1^2
						\notag \\
			=	&	\,	k \F{\Phi}_2^2 
						+ \frac{k^2}{2}  \F{\xi}_2 \F{\Phi}_1^2 
						- \frac{k^3}{8} \F{\xi}_1^2  \F{\Phi}_1^2  ,
	\label{eq:nonlinear2:Hkin4}
\end{align}
where I inserted \Eqs{eq:nonlinear2:phi_aux_I} and \eq{eq:nonlinear2:phi_aux_II} in the last line.  Equations \eq{eq:nonlinear2:Hkin2} and \eq{eq:nonlinear2:Hkin4} are later substituted into \Eq{eq:nonlinear2:Ham}.

\subsection{Calculation of the magnetic Hamiltonian $\boldsymbol{H_{\rm B}}$}
\label{app:nonlinear2:HB}

To calculate $H_{\rm B}$ in \Eq{eq:basic:H_magnetic}, I shall first write the magnetic potential $\psi$ in terms of the surface perturbation $\xi$.  The magnetic potential must satisfy the boundary condition \eq{eq:basic:boundary}.  Hence, the equations satisfied by the first two Fourier harmonics $\smash{\F{\psi}_1}$ and $\smash{\F{\psi}_2}$ are
\begin{align}
	(\vec{k} \cdot \vec{B}_0) \, \F{\xi}_1 
		&	=	- k \F{\psi}_1 
					- \ep^2 k^2  \F{\xi}_1 \F{\psi}_2
					- \ep^2 \frac{k^2}{2} \F{\xi}_2 \F{\psi}_1 		\notag \\
		&		\quad
					-\frac{3\ep^2}{8} k^3 \F{\psi}_1 \F{\xi}_1^2 +...
		\label{eq:nonlinear2:boundary1} \\
	2 (\vec{k} \cdot \vec{B}_0) \, \F{\xi}_2 
		&	=	-2 k \F{\psi}_2 -  k^2\F{\xi}_1  \F{\psi}_1  + ...
		\label{eq:nonlinear2:boundary_2}	
\end{align}
To solve the equations above, I use the following asymptotic ansatz:
\begin{equation}
	\F{\psi}_n = \F{\psi}_n^{(0)} + \ep^2 \F{\psi}_n^{(1)} + ...
	\label{eq:nonlinear2:psi_asymptotic}
\end{equation}
I now substitute \Eq{eq:nonlinear2:psi_asymptotic} into \Eqs{eq:nonlinear2:boundary1} and \eq{eq:nonlinear2:boundary_2}.  To zeroth order in $\ep$, the equation for $\smash{\F{\psi}_1}$ gives
\begin{equation}
	\F{\psi}_1^{(0)} = - k^{-1} (\vec{k} \cdot \vec{B}_0) \, \F{\xi}_1,
	\label{eq:nonlinear2:psi_10}
\end{equation}
and the equation for the second harmonic leads to
\begin{align}
	\F{\psi}_2^{(0)} 
		& = - k^{-1} (\vec{k} \cdot \vec{B}_0 ) \, \F{\xi}_2 - k  \F{\psi}_1^{(0)} \F{\xi}_1 /2 \notag \\
		& = - k^{-1} (\vec{k} \cdot \vec{B}_0 ) \, \F{\xi}_2 + (\vec{k} \cdot \vec{B}_0) \F{\xi}_1^2 /2.
	\label{eq:nonlinear2:psi_20}
\end{align}
To the second order in $\ep$, I obtain
\pagebreak
\begin{align}
	\F{\psi}_1^{(2)} 
		&	=  -  k \F{\xi}_1 \F{\psi}_{2}^{(0)} 
				- \frac{k}{2}  \F{\xi}_2 \F{\psi}_1^{(0)}  
				- \frac{3}{8} k^2 \F{\xi}_1^2 \F{\psi}_1^{(0)} 
			 	\notag \\
		&	=  \frac{3}{2}(\vec{k} \cdot \vec{B}_0) \, \F{\xi}_1 \F{\xi}_2 
				-\frac{k}{8}  (\vec{k} \cdot \vec{B}_0)   \F{\xi}_1^3,
	\label{eq:nonlinear2:psi_12}
\end{align}
where I substituted \Eqs{eq:nonlinear2:psi_10} and \eq{eq:nonlinear2:psi_20}.  Further higher order corrections for $\smash{\F{\psi}_{1}}$ or $\smash{\F{\psi}_{2}}$ are not needed.

After obtaining the asymptotic expansions for $\smash{\F{\psi}_{1,2}}$ in \Eqs{eq:nonlinear2:psi_asymptotic}--\eq{eq:nonlinear2:psi_12}, I can now calculate $H_{\rm B}$ in \Eq{eq:linear:HB}.  In the Fourier representation, \Eq{eq:linear:HB} can be written as
\begin{widetext}
\begin{align}
	H_{\rm B} 	=  & 	-\frac{1}{d_x d_y} \frac{ 1 }{4\pi \rho }	
									\sum_{n\in \mathbb{Z}^+} \sum_{m\in \mathbb{Z}^+} \int  \dx \, 
									m(\vec{k}	 \cdot \vec{B}_0 ) 	 \,							
									\left(1+n k \xi + \frac{n^2}{2}k^2\xi^2 + ... \right) \F{\psi}_n \F{\xi}_m
									\sin(n \vec{k}\cdot \vec{x}) \sin(  m \vec{k}\cdot \vec{x} ).					
\end{align}
\end{widetext}
As before, I write $H_{\rm B}$ as an asymptotic series
\begin{equation}
	H_{\rm B} = \ep^2 H_{\rm B,2} + \ep^3 H_{\rm B,3} + \ep^4 H_{\rm B,4} + ...
\end{equation}
and calculate each term order by order in $\ep$.  To lowest order, I obtain
\begin{align}
	H_{\rm B,2} 	
		&	= 	- \frac{ (\vec{k}	 \cdot \vec{B}_0 )  }{8\pi \rho  }		\,  \F{\xi}_{1}  \F{\psi}_1^{(0)} 
			=	 \frac{( \vec{k}	 \cdot \vec{v}_A )^2}{2k}  \, \F{\xi}_{1}^2,
	\label{eq:nonlinear2:Hb2}
\end{align}
where I substituted $\F{\psi}_1^{(0)}$ in \Eq{eq:nonlinear2:psi_10} and $\vec{v}_A \doteq \vec{B}_0 / \sqrt{4\pi \rho}$ is the Alfvén velocity.  This agrees with result in \Eq{eq:linear:HB_approx}.  As before, $H_{\rm B,3}$ is zero.  Finally, the $\mc{O}(\ep^4)$ contribution to $H_{\rm B}$ is
\begin{align}
	H_{\rm B,4} 	
		=	& 	
				-\frac{ (\vec{k}	 \cdot \vec{B}_0 ) } {8\pi \rho }	 
				\bigg( 	2  \F{\xi}_2 \F{\psi}_2^{(0)} 
							+ \F{\xi}_1 \F{\psi}_1^{(2)}
							+ k \F{\xi}_1^2 \F{\psi}_2^{(0)}
							\notag \\
			&	\qquad \qquad \qquad
							+ \frac{k}{2} \, \F{\xi}_1 \F{\xi}_2 \F{\psi}_1^{(0)} 
							+ \frac{k^2}{8} \F{\xi}_1^3 \F{\psi}_1^{(0)} +....
				\bigg) 
			\notag \\
		=	& 	\, (\vec{k}	 \cdot \vec{v}_A )^2 
				\bigg( 		\frac{1}{k} \,  \F{\xi}_{2}^2
							-	\frac{1}{2} \, 	\F{\xi}_1^2 \F{\xi}_{2} 
							-   \frac{k}{8} \F{\xi}_{1}^4  
				\bigg) ,
	\label{eq:app_Hb:Hb4}
\end{align}
where I substituted \Eqs{eq:nonlinear2:psi_10}--\eq{eq:nonlinear2:psi_12}.

\bibliographystyle{apsrev}
\bibliography{single_mode_MRT,foot}

\begin{thebibliography}{75}
\bibitem[{foo({\natexlab{a}})}]{foot:psi}
\bibinfo{note}{The field $\psi$ is written in term of sine functions so that
  $\vec{B}_1 = \del \psi$ is in phase with the surface perturbations.}
\bibitem[{foo({\natexlab{b}})}]{foot:cylindrical}
\bibinfo{note}{This is only true for the case of spatially homogeneous
  background magnetic fields. In the cylindrical case, where the magnetic field
  depends on the radius, this term would lead to nonvanishing contributions.}
\bibitem[{foo({\natexlab{c}})}]{foot:exp}
\bibinfo{note}{For large enough times, terms proportional to
  $\exp(2\gamma_{\vec{k}} t)$ will dominate over terms proportional to
  $\exp(\gamma_{2\vec{k}} t)$. This is valid as long as $\gamma_{2\vec{k}}$ is
  not too small.}
\end{thebibliography}

\end{document}